\documentclass[11pt, a4paper]{article}

\usepackage[utf8]{inputenc}
\usepackage[english]{babel}
\usepackage{color}
\usepackage{graphicx}
\usepackage{doi}
\usepackage{multirow}
\usepackage{booktabs}
\usepackage{amsmath,amsthm}
\usepackage{amsfonts,amssymb}
\usepackage{natbib}
\usepackage{url}
\usepackage{grffile}			
\usepackage{a4wide}
\usepackage{hyperref}			

\title{Quasi-geostrophic modes in the Earth's fluid core with an outer stably stratified layer}

\author{J. Vidal$^{1,2}$ and N. Schaeffer$^{1,2}$ \\[0.5cm]
\small  $^1$ Univ. Grenoble Alpes, ISTerre, F-38041 Grenoble, France \\
\small  $^2$ CNRS, ISTerre, F-38041 Grenoble, France}

\begin{document}
\maketitle

\begin{abstract}

Seismic waves sensitive to the outermost part of the Earth's liquid core seem to be affected by a stably stratified layer at the core-mantle boundary.
Such a layer could have an observable signature in both long-term and short-term variations of the magnetic field of the Earth, which are used to probe the flow at the top of the core.
Indeed, with the recent SWARM mission, it seems reasonable to be able to identify waves propagating in the core with period of several months, which may play an important role in the large-scale dynamics.

In this paper, we characterize the influence of a stratified layer at the top of the core on deep quasi-geostrophic (Rossby) waves.
We compute numerically the quasi-geostrophic eigenmodes of a rapidly rotating spherical shell, with a stably stratified layer near the outer boundary.
Two simple models of stratification are taken into account, which are scaled with commonly adopted values of the Brunt-Väisälä frequency in the Earth's core.

In the absence of magnetic field, we find that both azimuthal wavelength and frequency of the eigenmodes control their penetration into the stratified layer: the higher the phase speed, the higher the permeability of the stratified layer to the wave motion.
We also show that the theory developed by \cite{Takehiro_2001strati} for thermal convection extends to the whole family of Rossby waves in the core.
Adding a magnetic field, the penetrative behaviour of the quasi-geostrophic modes (the so-called fast branch) is insensitive to the imposed magnetic field and only weakly sensitive to the precise shape of the stratification.

Based on these results, the large-scale and high frequency modes (1 to 2 month periods) may be detectable in the geomagnetic data measured at the Earth's surface, especially in the equatorial area where the modes can be trapped.
\end{abstract}


\section{Introduction}

The existence of a stably stratified layer at the top of the Earth's liquid core can have consequences on different aspects of the Earth's core dynamics.
Such a layer can have a chemical origin with the accumulation of light elements coming from the growth of the inner core \citep{fearn1981,Braginsky_2007strati,Gubbins_2013strati} or from chemical interactions with the mantle if the outer part of the core is under-saturated in oxygen and silicium \citep{Buffett_2010stratib}.
It can also have a thermal origin, if the heat flux at the core mantle-boundary is lower than the adiabatic heat flux in the core.
With their recent estimations of the thermal iron conductivity in the Earth's core, \citet{De_Koker_2012strati} and \citet{Pozzo_2012strati} claim that such a layer can be thermally stable at the top of the core.

The strength of the stratification in such a layer is characterized by the Brunt-Väisälä frequency
\begin{equation}
	N = \sqrt{- \frac{g}{\rho} \frac{\text{d}\rho}{\text{d}r}}, 
\end{equation}
where $\rho (r)$ is the density profile as a function of the radius, and $g$ the acceleration of gravity, which is considered constant in the thin layer \citep[e.g.][]{Crossley_1984waves}.
A stably stratified layer corresponds to the case $N^2 > 0$, and a convective region to $N^2 < 0$.
The comparison between $N$ and the planetary rotation rate $\Omega$ allows to distinguish two stratification regimes: weak stratification when $N < 2\Omega$ and strong stratification when $N > 2\Omega$.

The presence of such a stratified layer at the top of the outer core has been advocated by some seismologists to explain travel-time anomalies of seismic body waves sensitive to the outermost part of the Earth's liquid core.
It is however difficult to detect small density anomalies in this region, because body waves are also affected by the strong heterogeneities of the D'' layer at the base of the mantle \citep{souriau2007deep}.
Optimistic estimations give a thickness of the stratified layer between 50-100 km \citep{Lay_1990strati} and 300 km \citep{Helffrich_2010strati}, with a weak stratification corresponding to $N / 2 \Omega \sim 0.1$ in the latter case.

In a stratified layer, waves in which the Coriolis and buoyancy forces play an important role exist \citep{Friedlander_1982waves}.
These inertia-gravity waves have been investigated for several decades in the geophysical and astrophysical contexts \citep{Olson_1977waves,Crossley_1980waves,Crossley_1984waves,Friedlander_1985waves,Dintrans_1999}.
If we could detect them, they would also provide evidence for the existence of a stably stratified layer in the outermost part of the Earth's core. 
The possible detection of these waves in the geodetic high frequency spectrum attracted attention some years ago, thanks to the development of superconducting gravimeters \citep{Melchior_1986waves,Aldridge_1987waves}.
Unfortunately, the observations in the high frequency range do not require a stably stratified layer to explain the data \citep{Melchior_1988waves}, mainly because of the uncertainty on the frequencies of the waves \citep{Rieutord_1995waves}.
The evolution of the magnetic field of the Earth also gives constraints on the stratification at the top of the core \citep{Gubbins_2007strati}.

Although still debated, and with characteristics such as depth and $N$ not known, the existence of such a layer would give constraints on the thermal history of the core \citep{Labrosse_1997strati,Lister_1998strati,labrosse2003thermal}.
This layer may also modify the dynamics of the Earth's liquid core.
\citet{Zhang_1997strati}, \citet{Takehiro_2001strati,Takehiro_2002strati} and more recently \citet{Nakagawa_2011strati} studied the penetration of columnar convective motions of a rapidly rotating spherical shell in a stably stratified layer located at the outer boundary.
Their results show that the stratification acts as an interface, which can be crossed by large-scale motions while small-scale ones are trapped below the layer when the stratification is strong enough.
Such a layer has then the ability to partially hide geophysical flows inside the Earth's core, which has important implications for core flow inversion.

Indeed, using recordings of the magnetic field, it is possible to infer the flow at the core surface using inversion techniques \citep{holme07}.
Assuming a quasi-geostrophic (columnar) flow, it is then possible to reconstruct the flow in the core interior \citep[see e.g.][]{pais08}.
Although there is evidence for quasi-geostrophic dynamics in the Earth's core \citep{gillet2011}, the presence of a stratified layer may prevent us to reconstruct part of the interior flow \citep{bloxham1990consequences}.
Moreover, a stratified layer can host its own eigenmodes, which may then be distinct from the flow in the bulk of the core \citep{braginsky1998magnetic, braginsky1999dynamics, Buffett_2014strati}.

In this study, we characterize the effect of a stratified layer on otherwise quasi-geostrophic inertial modes in the Earth's core, also known as Rossby modes. Their time-scale is typically of a few months, and their signature may be actually seen in the geomagnetic data from the new SWARM mission \citep{swarm}.
Using a numerical eigenmode solver, we compute the linear hydromagnetic modes in a simplified model of the Earth's core including a stratified layer at the outer boundary.
The effect of an imposed magnetic field on the modes is briefly considered in the case of a simple toroidal magnetic field depending only on the cylindrical radius $s$ \citep{Malkus_1967waves}.

The paper is divided as follows.
Section \ref{sec:math} presents the mathematical modelling and section \ref{sec:num} the numerical methods used to solve the problem.
The results are shown in section \ref{sec:results}, with an emphasis on the shape of the eigenmodes and the dependence of their penetration length on governing parameters.
Section \ref{sec:discussion} discusses our findings.
We end the paper with a conclusion.

\section{Mathematical modelling}	\label{sec:math}

\subsection{Basic equations}
\begin{figure}
	\centering
	\begin{tabular}{cc}
	\includegraphics[width=0.4\textwidth]{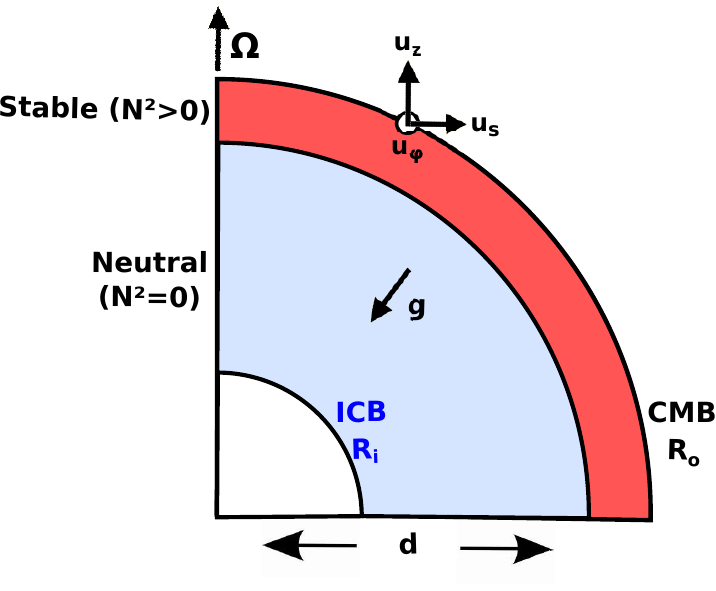} & \includegraphics[width=0.48\textwidth]{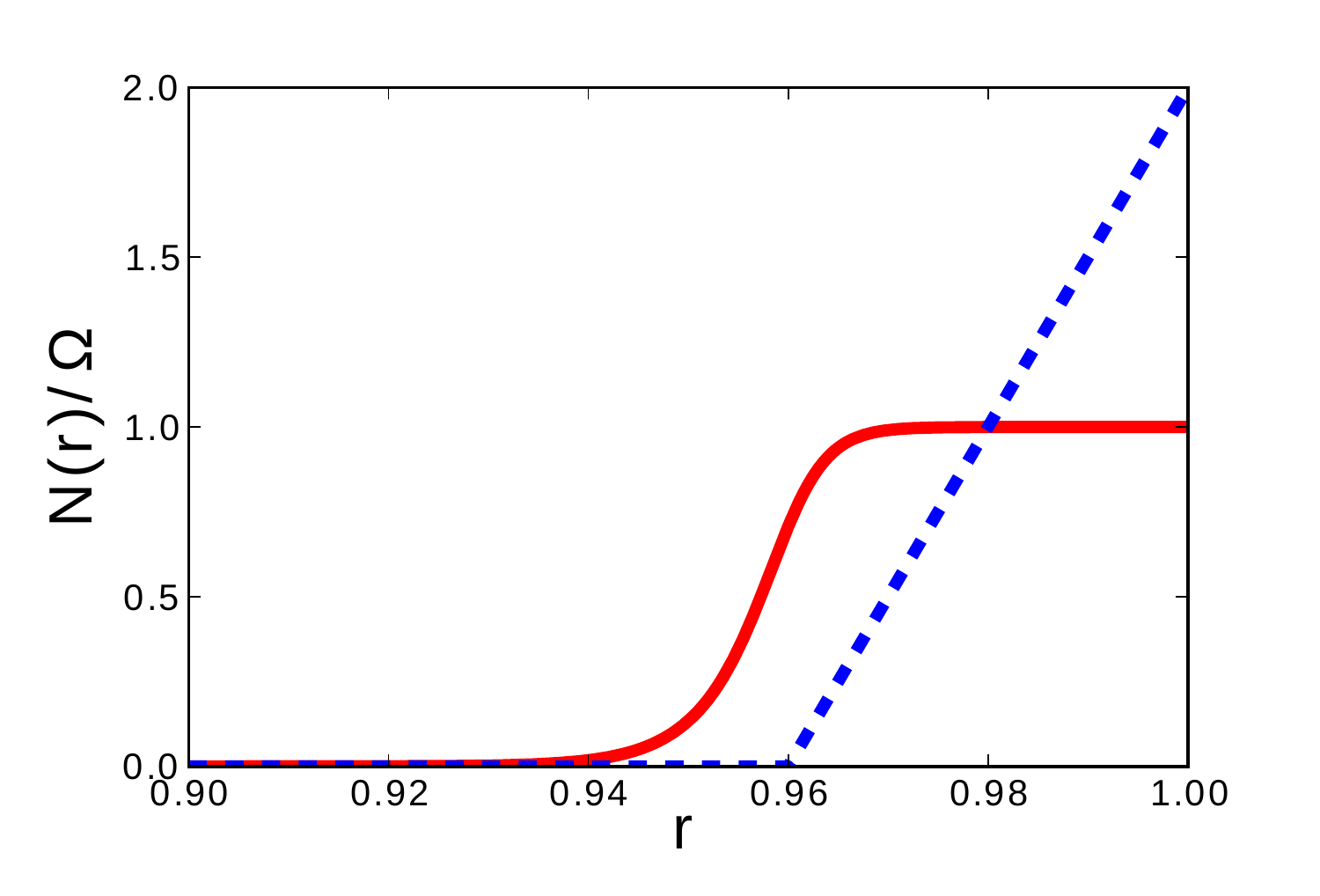}\\
	(a) & (b) \\
	\end{tabular}
	\caption{(a) Geometry of the system. (b) Dimensionless Brunt-Väisälä frequency $N (r)/\Omega$ as a function of the dimensionless radius $r$. The solid red and dashed blue curves correspond to model (\ref{Eq_Model_Takehiro}) and model (\ref{Eq_Model_Linear_Profile}) respectively, with $N_0 = \Omega$. The thickness of the stratified layer is 140 km.}
	\label{Fig_Model_Geometry}
\end{figure}

The geometry of the system is illustrated in the Figure \ref{Fig_Model_Geometry}a.
We consider a rapidly rotating spherical shell of inner radius $R_i$ and outer radius $R_o$, surrounded by perfectly insulating mantle and inner core.
We assume an aspect ratio $\gamma = R_i / R_o = 0.35$, similar to that of the Earth's core.
The shell, the mantle and the inner sphere are rotating at the same angular velocity $\mathbf{\Omega} = \Omega \, \mathbf{\hat{z}}$ aligned with the unitary vertical axis $\mathbf{\hat{z}}$ in cylindrical polar coordinates ($s,\phi, z$). We work in the rotating reference frame attached to the mantle, in cylindrical polar coordinates or in spherical polar coordinates $(r,\theta, \phi)$. The velocity $\mathbf{v}$, the magnetic field $\mathbf{B}$, the temperature $T$ and the density $\rho$ are subjected to small perturbations around a basic state $(\mathbf{U_0}, \mathbf{B_0}, T_0,\rho_0)$. Let us write
\begin{align}
	\left [ \mathbf{v}, \mathbf{B} \right ] (\mathbf{r},t) &= \left [ \mathbf{U_0}, \mathbf{B_0} \right ] (\mathbf{r}) + \left [ \mathbf{u}, \mathbf{b} \right ] (\mathbf{r},t), \\
	T (\mathbf{r},t) &= T_0 (r) + \Theta (\mathbf{r},t), \\
	\rho (\mathbf{r},t) &= \rho_0 \left ( 1 - \alpha \Theta (\mathbf{r},t) \right ),
\end{align}
where $\alpha$ is the volume coefficient of thermal expansion and $(\mathbf{u}, \mathbf{b}, \Theta, -\rho_0 \alpha \Theta)$ the perturbations of small amplitude. For the sake of simplicity, we impose $\mathbf{U_0} (\mathbf{r}) = \mathbf{0}$, which corresponds to a solid-body rotation in the inertial frame. We also assume that the background density $\rho_0$ is constant, neglecting the variations due to adiabatic compression across the shell \citep{Rieutord_1995waves}. We take into account the density perturbation only in the gravity term with the Boussinesq approximation, considering a gravity field of the form $\mathbf{g} = - g_o r \, \mathbf{\hat{r}}$ with $g_o R_o$ the acceleration of gravity at the outer boundary and $\mathbf{\hat{r}}$ the unitary radial vector. 

To keep the model simple, we suppose that the background magnetic field $\mathbf{B_0}$ in the shell is either an axial field 
\begin{equation}
	\mathbf{B_0} = B_0 \, \boldsymbol{\hat{z}},
	\label{Eq_Model_Bz}
\end{equation}
or the Malkus field \citep{Malkus_1967waves}
\begin{equation}
	\mathbf{B_0} (\mathbf{r}) = B_0 \, \mathbf{b_0}(\mathbf{r}) = B_0 \, r \sin \theta \, \boldsymbol{\hat{\phi}}.
	\label{Eq_Model_Malkus}
\end{equation}
The latter is a purely toroidal and azimuthal field which increases with the cylindrical radius $s=r \sin \theta$. This assumption is consistent with our choice for the background velocity, since no basic velocity $\mathbf{U_0}$ is needed for the basic state to be in equilibrium (the resulting Lorentz force is balanced by the pressure gradient).


We use $R_o$ as the length scale, $\Omega^{-1}$ as the time scale, $\rho_0 R_o \Omega^2$ as the pressure scale, $\Omega^2 R_o / \alpha g_o$ as the temperature scale and $\sqrt{\mu_0 \rho_0} \, \Omega R_o$ as the magnetic scale. As we are interested by the eigenmodes, we neglect the non-linear terms and the dimensionless equations satisfied by the perturbations are

\begin{align}
	\frac{\partial \mathbf{u}}{\partial t} + 2 \, \mathbf{\hat{z} \times \mathbf{u}} &= - \nabla P + Le \, \left [ (\nabla \times \mathbf{b}) \times \mathbf{b_0} + (\nabla \times \mathbf{b_0}) \times \mathbf{b} \right ] + r \Theta \, \mathbf{\hat{r}} + E \, \nabla^2 \mathbf{u} \label{Eq_Model_NS} \\
	\frac{\partial \Theta}{\partial t} &= -\frac{N^2 (r)}{\Omega^2} \, \mathbf{u} \cdot \mathbf{\hat{r}} + \frac{E}{Pr} \nabla^2 \Theta, \label{Eq_Model_Temp} \\
	\frac{\partial \mathbf{b}}{\partial t} &= Le \, \nabla \times \left ( \mathbf{u} \times \mathbf{b_0} \right ) + E_\eta \, \nabla^2 \mathbf{b}. \label{Eq_Model_Mag}
\end{align}
Five governing parameters appear in the previous equations, namely the Ekman number $E$, the magnetic Ekman number $E_\eta$, the Prandtl number $Pr$, the Lehnert number $Le$ and the dimensional Brunt-Väisälä frequency $N$. They are defined by
\begin{equation}
	E = \frac{\nu}{\Omega R_o^2}, \, \ E_\eta = \frac{\eta}{\Omega R_o^2}, \, \ Pr = \frac{\nu}{\kappa}, \, \ Le = \frac{B_0}{\sqrt{\rho_0 \mu_0}\, \Omega R_o} \, \ \text{and} \, \ N(r) = \sqrt{\alpha g_o \frac{\text{d} T_0}{\text{d}r}},
\end{equation}
with $\nu$, $\kappa$ and $\eta$ respectively the viscous, thermal and magnetic diffusivities and $B_0$ the strength of the background magnetic field. In the computations, we take $Pr=0.1 - 10$ and $Le=10^{-4}$. Because of numerical constraints, we choose $E=E_\eta =10^{-7}$.

Equations (\ref{Eq_Model_NS}), (\ref{Eq_Model_Temp}) and (\ref{Eq_Model_Mag}) are completed with boundary conditions on the velocity, the temperature and the magnetic field. We impose on both shells the no-slip boundary condition for the velocity field 
\begin{equation}
	\mathbf{u} = \mathbf{0}
\end{equation}
and a constant heat flux
\begin{equation}
	\frac{\partial \Theta}{\partial r} = 0.
\end{equation}
The magnetic field boundary conditions are the continuity of $\mathbf{b}$ and the continuity of the radial component of $\nabla \times \mathbf{b}$.
The spherical harmonic expansion (introduced below) allows to write this as a boundary condition.

\subsection{Basic state for the stratification}

The Brunt-Väisälä frequency $N(r)$ in the core is highly speculative. Following \citet{Lister_1998strati}, let us assume that the spherical shell is divided in two parts (Figure \ref{Fig_Model_Geometry}a).
A stably stratified layer with $N^2 (r)>0$ is located above a convective core of dimensionless radius $d$, which is supposed to be neutral with respect to the stratification ($N^2 (r) = 0$).
This assumption filters out the coupling between waves and convection \citep{Rieutord_1995waves}.
We set the radius $d=0.96$, which corresponds to a stratified layer of thickness 140 -150 km. This thickness belongs to the range estimated by seismic studies \citep{Helffrich_2010strati}. It is also inferred by \citet{Buffett_2014strati} to match the geomagnetic secular variation of the axial dipole with slow Magneto-Archimedes-Coriolis (MAC) waves.
Following \citet{Takehiro_2001strati,Takehiro_2002strati} and \citet{Nakagawa_2011strati}, we adopt
\begin{equation}
	N (r) = N_0\sqrt{1-\frac{1}{2} \left [ 1 - \tanh \left ( \frac{r - d}{a} \right ) \right ]}
	\label{Eq_Model_Takehiro}
\end{equation}
which is illustrated in Figure \ref{Fig_Model_Geometry}b.
$N_0$ is the Brunt-Väisäla frequency at the outer boundary and $a=0.005$ the transition thickness between the neutral core and the stratified layer.
$N(r)$ is thus a smooth profile in the core with a roughly constant value $N(r) \simeq N_0$ in the outer layer.
Different values of $N_0$ are taken into account.
In addition, we also consider a profile similar to the one used by \citet{Rieutord_1995waves} and \citet{Buffett_2014strati}, \emph{i.e.}
\begin{equation}
	N(r) = \left \{
	\begin{array}{ll}
		2 N_0 (r - d)/(1 - d) & \forall \ r \geq d \\
		0 & \forall \ r < d \\
	\end{array}
	\right .
	\label{Eq_Model_Linear_Profile}.
\end{equation}
It is linear in the stratified layer, with a mean value $N_0$ and a maximum value $2N_0$ at the outer boundary.

\section{Numerical modelling}	\label{sec:num}

\subsection{Spectral method}
As they are solenoidal, $\mathbf{u}$ and $\mathbf{b}$ are decomposed into poloidal and toroidal parts
\begin{align}
	\mathbf{u} &= \nabla \times \nabla \times (U \mathbf{r}) + \nabla \times (V \mathbf{r}), \\
	\mathbf{b} &= \nabla \times \nabla \times (G \mathbf{r}) + \nabla \times (H \mathbf{r}),
\end{align}
with $(U,G)$ and $(V,H)$ the poloidal and toroidal scalars. The scalars $(U,V,G,H,\Theta)$ are then expanded on orthonormalized spherical harmonics $Y_l^m (\theta, \phi)$, of degree $l$ and order $m$, as
\begin{align}
	\left [ U, V, \Theta \right ] (\mathbf{r},t) &= \sum_{l=m}^{\infty} \left [ U_l^m, V_l^m, \Theta_l^m \right ] (r) \, Y_l^m (\theta, \phi) \exp(\lambda t), \label{Eq_Model_Num_SH_V} \\
	\left [ G, H, \right ] (\mathbf{r},t) &= \sum_{l=m}^{\infty} \left [ G_l^m, H_l^m \right ] (r) \, Y_l^m (\theta, \phi) \exp(\lambda t).
	\label{Eq_Model_Num_SH_B}
\end{align}
Since the basic state $(\mathbf{U_0}, \mathbf{B_0})$ is axisymmetric, there is no coupling between Fourier modes $m$ in the equations
(\ref{Eq_Model_NS}) to (\ref{Eq_Model_Mag}).
Thus, one can seek solutions $(\mathbf{u}, \Theta, \mathbf{b})$ with a single azimuthal wavenumber $m$.
As we are interested in the eigenmodes of the system, we assume that the time dependence of all perturbations is proportional to $\exp(\lambda t)$, with the complex eigenvalue
\begin{equation}
	\lambda = \tau + i \omega
\end{equation}
where $\tau$ is the growth rate of the mode and $\omega$ its frequency, both non-dimensional.

Thanks to the spectral decomposition, the boundary conditions are easy to express.
On the two boundaries, the no-slip condition becomes
\begin{equation}
	U_l^m = \frac{\text{d} U_l^m}{\text{d} r} = V_l^m = 0
\end{equation}
and the constant heat flux condition
\begin{equation}
	\frac{\text{d} \Theta_l^m}{\text{d} r} = 0.
\end{equation}
With the insulating boundary condition, the magnetic field matches a potential field in both the mantle and the inner sphere, resulting to
\begin{equation}
	\frac{\text{d} G_l^m}{\text{d}r} - \frac{l}{r} G_l^m = H_l^m = 0
\end{equation}
on the inner boundary ($r = \gamma$) and
\begin{equation}
	\frac{\text{d} G_l^m}{\text{d}r} + \frac{l+1}{r} G_l^m = H_l^m = 0
\end{equation}
on the outer boundary ($r=1$).

\subsection{Symmetry considerations}

The equations (\ref{Eq_Model_NS} - \ref{Eq_Model_Mag}) are symmetric or anti-symmetric with respect to the equatorial plane.
This implies that solutions can be characterized by their symmetry with respect to the equator \citep{Gubbins_1993}.

In the following, we focus on equatorially symmetric $\mathbf{u}$ and $\Theta$, because quasi-geostrophic flows have this equatorial symmetry.
They satisfy
\begin{align}
	[u_r, u_\theta, u_\phi] (r, \theta, \phi) &= [u_r, -u_\theta, u_\phi]  (r, \pi - \theta, \phi), \\
	\Theta (r, \theta, \phi) &= \Theta (r, \pi - \theta, \phi).
\end{align}
Taking the symmetry of $Y_l^m$ into account, the spherical harmonic expansions for the poloidal and toroidal components of $\mathbf{u}$ (and $\mathbf{b}$) alternate to match the equatorial symmetry.
Note also that the Coriolis force couples $U_l^m$ and $V_{l\pm 1}^m$.
Following \citet{Schmitt_2010}, the symmetric part of $\mathbf{u}$ and $\Theta$ is given by
\begin{equation}
	(U_l^m, V_{l+1}^m, \Theta_l^m),
\end{equation}
for a given azimuthal number $m$ and with $l=m,m+2,m+4,\dots$

Because $\mathbf{u}$ is symmetric, the symmetry of $\mathbf{b}$ is the same as the one of $\mathbf{B_0}$:
$\mathbf{b}$ is equatorially symmetric for the Malkus field, while it is equatorially antisymmetric for the axial magnetic field.
Hence, for the magnetic field we add to the state vector
\begin{equation}
	(G_l^m, H_{l+1}^m) \quad \text{or} \quad (G_{l+1}^m, H_l^m)
\end{equation}
respectively for symmetric or antisymmetric imposed fields.

\subsection{Numerical implementation}
\begin{figure}[t]
	\centering
	\includegraphics[width=0.65\textwidth]{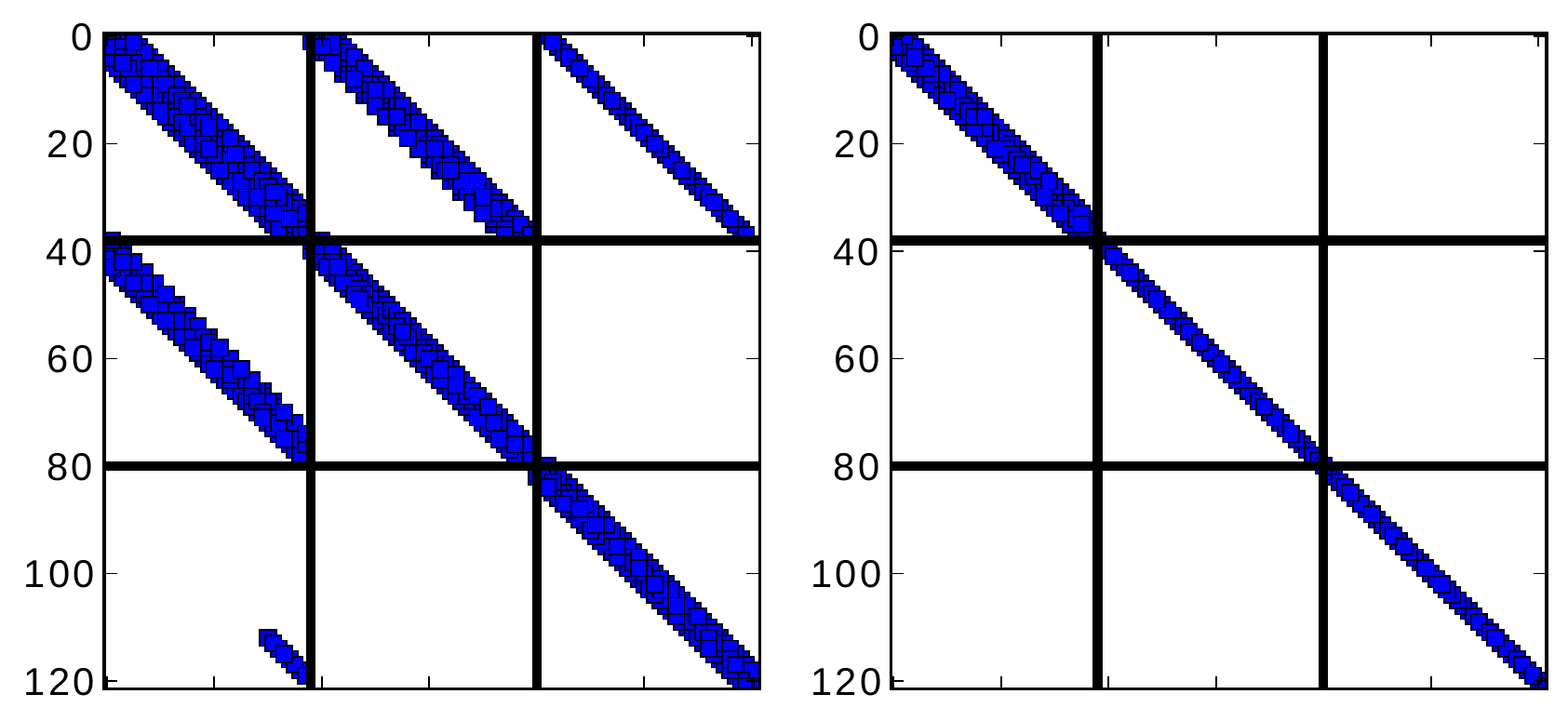}
	\caption{Sparse matrices $A$ and $B$ in the non-magnetic case. $L_{max} = 6$ and $N_r = 20$.
	The black lines separate the poloidal and toroidal components of the velocity and the temperature field.
	The first line of blocks represents the evolution equation of the poloidal component of the velocity (coupled to the toroidal flow by the Coriolis force and to the temperature by the Buoyancy force), the second line is the equation for the toroidal velocity, and the last line is the equation for the temperature (coupled to the poloidal flow only in the stratified layer).}
	\label{Fig_Model_Num_Matrix_A}
\end{figure}

Substituting from (\ref{Eq_Model_Num_SH_V}) and (\ref{Eq_Model_Num_SH_B}) into the governing equations (\ref{Eq_Model_NS}) to (\ref{Eq_Model_Mag}) leads to a set of linear differential equations which forms a generalized eigenvalue problem with its boundary conditions.
To get numerical solutions, the spherical harmonic expansions are truncated at the harmonic degree $L_{M}$ and the differential operators are represented by a second order finite difference scheme on an irregular mesh of $N_r$ levels, adjusted to properly resolve the Ekman boundary layers.
The eigenvalue problem is now in a matrix form as a complex, non-hermitian and generalized eigenvalue problem 
\begin{equation}
	A \mathbf{X} = \lambda B \mathbf{X}.
	\label{Eq_Model_Eigen}
\end{equation}
$\lambda$ is the eigenvalue and $\mathbf{X}$ is the associated symmetric eigenvector given by
\begin{equation}
	\mathbf{X} = \left ( U_l^m (r_i), V_{l+1}^m (r_i), \Theta_l^m (r_i), G_l^m (r_i), H_{l+1}^m (r_i) \right )^t,
\end{equation}
with $i=0,1,\dots,N_r$ and $l=m,m+2,m+4,\dots$ for a symmetric eigenmode permeated by a symmetric imposed field.
The two matrices $A$ and $B$, collecting the terms on the left-hand side and right-hand side of the equations, have approximately $(5 L_{M} N_r /2)^2$ elements.
In order to have good spatial and spectral resolutions, we use $L_{M} \sim 350$ and $N_r \simeq 850$ for Ekman number $E=10^{-7}$.
Matrices $A$ and $B$ have a large number of elements, but most of them are zero.
They are stored as sparse matrices in order to reduce the required memory (see Figure \ref{Fig_Model_Num_Matrix_A}).

Following \cite{Rieutord_1997} and \cite{Dintrans_1999}, the linear eigenvalue problem (\ref{Eq_Model_Eigen}) is solved using the shift-and-invert spectral transformation: instead of solving the problem (\ref{Eq_Model_Eigen}), the related problem 
\begin{equation}
	(A - \sigma B)^{-1} B \mathbf{X} = \mu \mathbf{X}
\end{equation}
is solved with $\sigma$ the shift and the new eigenvalue $\mu$ related to $\lambda$ by
\begin{equation}
	\mu = \frac{1}{\lambda - \sigma}.
\end{equation}
With this spectral transformation, it is easier to get the least damped quasi-geostrophic modes, imposing $\sigma = i \omega$ with $\omega \ll 1$. Only eigenvalues with absolute residuals $|(A - \sigma B)^{-1} B \mathbf{X} - \mu \mathbf{X}| < 10^{-12}$ are accepted as good eigenvalues.

We wrote a Python code, nicknamed \emph{Singe} (Spherical INertia-Gravity Eigenmodes), that relies on the \texttt{SLEPc} library \citep{Hernandez_2005slepc} to solve the eigenvalue problem using the restarted Krylov-Schur method \citep{SLEPc_STR7} together with the shift-and-invert strategy.
\texttt{SLEPc} is an open-source scientific library developed to efficiently solve large and sparse eigenvalue problems on parallel computers.
It is itself based on \texttt{PETSc} \citep{petsc-efficient,petsc-web-page}, another open-source library intended to solve scientific problems modelled by partial differential equations.
Finally, \emph{Singe} uses the \texttt{SHTns} library \citep{Schaeffer_2013} for the spherical harmonic transforms.

The \emph{Singe} code was benchmarked with the theoretical solutions of \citet{zhang2001inertial}, \citet{liao2001viscous} and \citet{zhang2003nonaxisymmetric} in the full sphere.
As benchmarks for the spherical shell geometry, we used the numerical results of \citet{Rieutord_1997} without stratification and of \citet{Dintrans_1999} with a constant stratification in the shell.

\emph{Singe} is available at \url{https://bitbucket.org/nschaeff/singe} as free software.

\section{Results}	\label{sec:results}

\begin{table}[t]
	\centering
	\begin{tabular}{cccccccc}
	\toprule
	\multirow{2}{*}{$m$} & \multirow{2}{*}{$n_c$} & \multicolumn{6}{c}{$\omega$} \\
	{} & {} & $N_0 = 0$ & $N_0=\Omega$ & $N_0=2 \, \Omega$ & $N_0=10\,\Omega$ & solid SF & solid NS \\
	\toprule
	\multirow{2}{*}{1} & 8 & $-0.0046$ & $-0.0051$ & $-0.0054$ & $-0.0060$ & $-0.0051$ & $-0.0043$\\
		{} & 3 & $-0.014$ & $-0.017$ & $-0.020$ & $-0.023$ & $ -0.015$ & $-0.014$ \\
	\multirow{2}{*}{3} & 13 & $-0.0039$ & $-0.0044$ & $-0.0048$ & $-0.0051$ & $-0.0042$ & $-0.0036$\\
	{} & 7 & $-0.019$ & $-0.020$ & $-0.023$ &  $-0.025$ & $-0.020$ & $-0.019$ \\
	\multirow{2}{*}{6} & 13 & $-0.0069$ & $-0.0075$ & $-0.0080$ & $-0.0084$ & $-0.0072$ & $-0.0066$\\
	{} & 8 & $-0.021$ & $-0.024$ & $-0.029$  & $-0.031$ & $-0.022$ & $-0.021$ \\
	\multirow{2}{*}{8} & 14 & $-0.0079$ & $-0.0084$ & $-0.0093$ & $-0.0093$ & $-0.0083$ & $-0.0076$ \\
	{} & 10 & $-0.018$ & $-0.020$ & $-0.022$ & $-0.024$ & $-0.018$ & $-0.017$ \\
	\bottomrule
	\end{tabular}
	\caption{Wave number $m$ (azimutal), approximative number of quasi-geostrophic columns along a meridian $n_c$, dimensionless frequency $\omega$ and period in months of some numerical eigenmodes for four values of $N_0$, without magnetic field.
	Computations at $E=10^{-7}$ and $Pr=0.1$. $N(r)$ is given by (\ref{Eq_Model_Takehiro}).
	In addition, we have also computed the same modes by replacing the stratified layer by a solid layer, with stress-free (solid SF) or no-slip (solid NS) boundary conditions.
	}
	\label{Table_Penetration_Eigenmodes}
\end{table}

\subsection{Description of eigenmodes}

To study the behaviour of quasi-geostrophic inertial modes in the presence of a stratified layer, we first focus on the non-magnetic case.
Characteristics of some modes are gathered in Table \ref{Table_Penetration_Eigenmodes}.
Each mode is specified by a triplet $(\omega,m,n_c)$, with $m$ the azimuthal wave number and $n_c$ the number of zero-crossings of a velocity component ($u_s$ or $u_\phi$) along a cylindrical radial profile (measuring latitudinal variations).
The dimensionless frequency $\omega$, with corresponding periods ranging from one to several months, is always negative because the eigenmodes propagate eastwards.
It is a characteristic of Rossby waves in the Earth's core geometry \citep[e.g.][]{busse1970}.
Indeed, looking at the different forces, we find that the modes are dominated by the Coriolis force, with some exceptions in the stably stratified layer (see below).
The Coriolis force tends to align the flow along the spin axis to create Taylor columns, and the modes are mainly quasi-geostrophic inertial modes (almost invariant along the rotation axis).
The frequency depends also on the length scale of the modes.
For a given azimuthal number $m$, high frequency modes are large scale modes with a small number of columns $n_c$, while low frequency are small scale modes with a large number of columns.

The precise shape of the modes depends on the stratification, which may influence the penetration of the flow in the outer layer \citep{Zhang_1997strati,Takehiro_2001strati}.
However, the overall pattern of the modes in the neutral part of the core smoothly evolves with the stratification strength $N_0$.
For commonly adopted values of the Brunt-Väisälä frequency $N_0$, the number of columns of a mode remains approximatively the same, while the exact position and thickness of the columns change near the outer layer. 
A smooth increase of $N_0$ allows us to study the evolution of the modes.
It can be done with an iterative process to follow a given mode defined by $(m,n_c)$.
As observed in Table \ref{Table_Penetration_Eigenmodes}, $\omega$ slowly increases with $N_0$.
This slow increase of the frequency $\omega$ cannot be attributed to a geometrical effect.
Indeed, when replacing the stratified layer by a solid layer, the variation in frequency is much smaller, especially at large wave-numbers (see Table \ref{Table_Penetration_Eigenmodes}).

\begin{figure}
	\centering
	\includegraphics[width=1\textwidth]{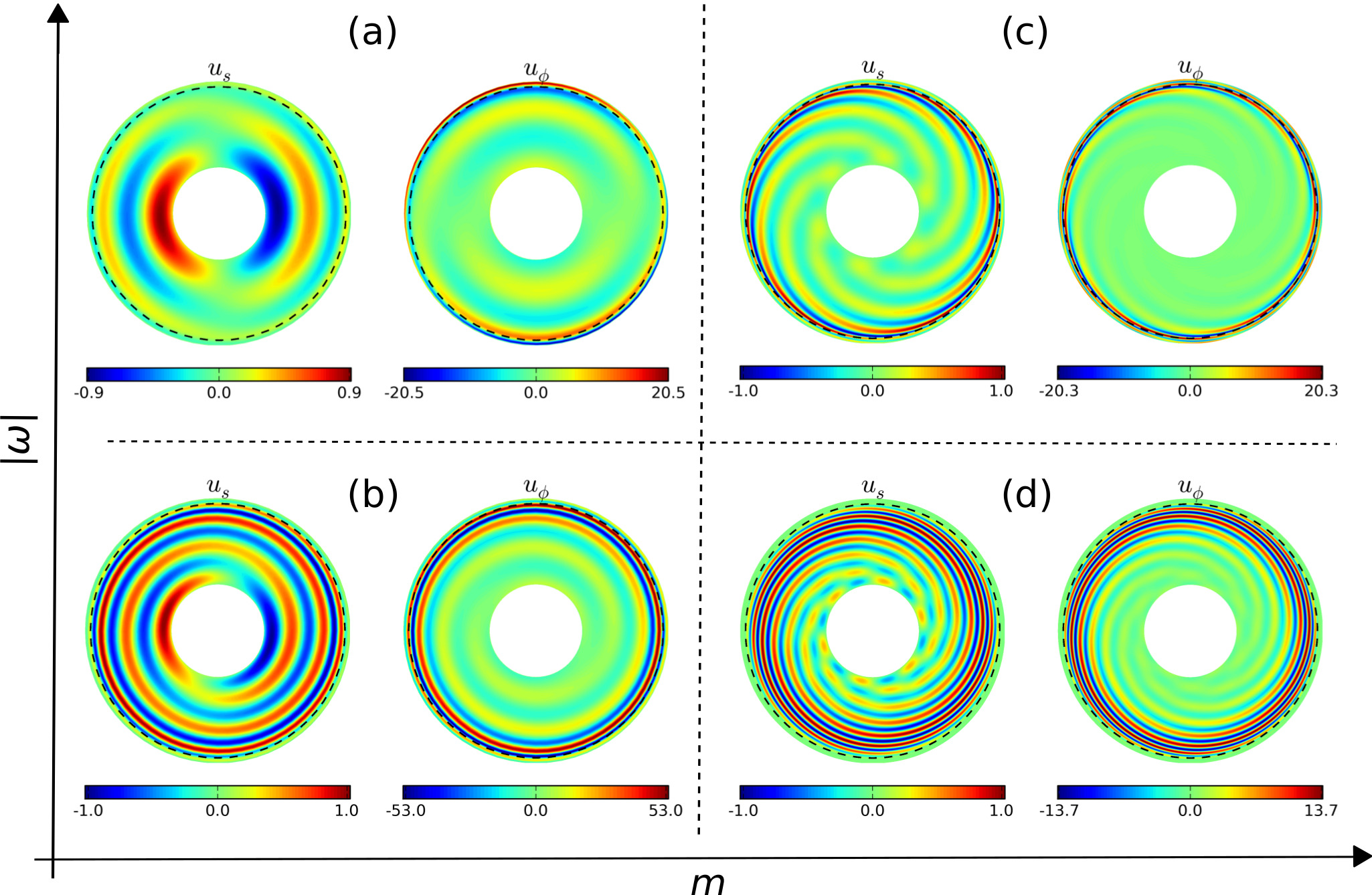}
	\caption{Contours of the cylindrical components $u_s$ and $u_\phi$ of four numerical eigenmodes from Table \ref{Table_Penetration_Eigenmodes} (without magnetic field) in the plane ($m, |\omega|$). $\omega$ is the dimensionless frequency.
	Contours are shown in the equatorial plane as seen from above. The vertical component $u_z$ is not shown because it vanishes in the equatorial plane.
	The velocity is normalized by the maximum value of $u_s$. The black dashed circles represent the transition between the neutral core and the stably stratified layer.
	The modes were computed at $E=10^{-7}$ and $Pr=0.1$.
	$N(r)$ is given by (\ref{Eq_Model_Takehiro}) and $N_0=\Omega$. (a) High-frequency mode $m=1$. (b) Low-frequency mode $m=1$. (c) High-frequency mode $m=6$. (d) Low-frequency mode $m=6$.}
	\label{Fig_Penetration_Eigenmodes}
\end{figure}

The cylindrical components $u_s$ and $u_\phi$ in the equatorial plane of some modes are shown in Figure \ref{Fig_Penetration_Eigenmodes}.
It is worth noting that lower frequency modes are spiralling more than higher frequency modes.
The spiralling appears because the viscosity is overestimated in our computations (see appendix \ref{sec:visc} for more details).
Finally, the kinetic energy is dominated by the azimuthal component of the velocity $u_\phi$, which is more concentrated in the equatorial region when the frequency is higher.
This phenomenon, called trapping, is known for inertial waves \citep{zhang1993equatorially} as well as for inertia-gravity waves \citep{Crossley_1984waves,Friedlander_1985waves}.
When $E < 10^{-7}$, we observe that this trapping becomes more efficient.
Unfortunately, it is then expensive to resolve numerically, because it needs enough grid points to capture the boundary layer and enough spherical harmonics to describe the trapping.
For these reasons, we set the Ekman number to $10^{-7}$ in our computations.

Let us now switch on the background magnetic field $\mathbf{B_0}$ given by expression (\ref{Eq_Model_Malkus}).
Without stratification, a background magnetic field splits the modes into "slow" and "fast" magneto-Coriolis (MC) waves, the latter being essentially pure inertial waves \cite[see e.g.][]{zhang2003nonaxisymmetric}.

We computed these fast modes with $Pr=0.1$ and a Lehnert number $Le = 10^{-4}$ corresponding to $B_0 = 3\,$mT \citep[in the range estimated for the Earth's core, see][]{gillet2010}.
For the Malkus field (eq. \ref{Eq_Model_Malkus}), we were able to compute the modes down
to $E = E_\eta = 10^{-7}$.
Computation of the modes with an axial field (eq. \ref{Eq_Model_Bz}) was numerically more demanding, probably because of the additional coupling terms in the corresponding matrix.
Indeed, in that case we were able to compute the modes down to $E = E_\eta = 10^{-5}$ only.
For azimuthal wave-numbers ranging from $m=1$ to $m=26$, the frequency $\omega$ changes by less than 1\% and the shape of the modes is indistinguishable from the ones obtained without magnetic field.
Note however that the overall damping rate of the mode is affected, especially when $E=E_\eta=10^{-5}$, but we expect it to remain small in the Earth's core, where $E_\eta \sim 10^{-10}$.
This holds for every eigenmode considered here, for all values of the Brunt-Väisälä frequency $N_0$ investigated here, and for the two magnetic field models (Malkus and axial).
It shows that the eigenmodes considered in this study are barely modified by the background magnetic field, because they are mainly "fast" quasi-geostrophic inertial modes.
These results are in agreement with the theoretical predictions of \citet{Malkus_1967waves} and \citet{zhang2003nonaxisymmetric} for a homogeneous fluid contained in a full sphere.
Furthermore, \citet{Canet_2014} considered a quasi-geostrophic model of magneto-Coriolis (MC) waves and showed that various shapes of a toroidal background magnetic fields do not affect the fast modes either.

Since our quasi-geostrophic modes are not affected significantly by the magnetic field, with or without a stratified layer, we will no longer consider
the magnetic field in the remainder of this paper, effectively solving only equations (\ref{Eq_Model_NS}) and (\ref{Eq_Model_Temp}) with $Le=0$.
This simplifies the numerical study quite a bit.
Note that although the fast branch considered here is unaffected, the slow branch will most certainly be affected by the magnetic field, but this is beyond the scope of this paper.

\begin{figure}
	\centering
	\includegraphics[width=0.95\textwidth]{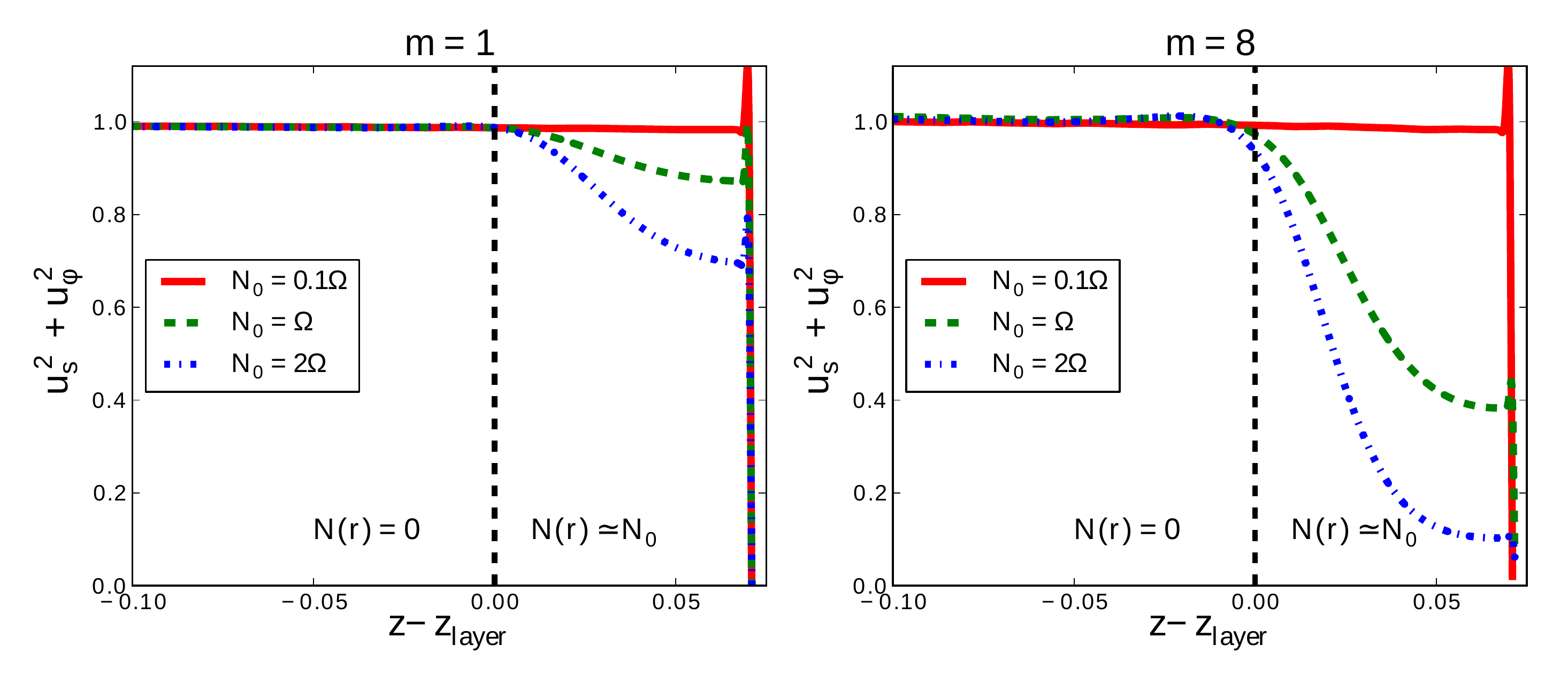}
	\caption{Distribution of the azimuthal average of the quasi-geostrophic kinetic energy $u_s^2 + u_\phi^2$, along a quasi-geostrophic column located near the cylindrical radius $s=0.7$, for some high frequency modes of the Table \ref{Table_Penetration_Eigenmodes}, computed at $E = 10^{-7}$ and $Pr=0.1$. $N(r)$ is given by eq. (\ref{Eq_Model_Takehiro}).
	The horizontal axis $z-z_{layer}$ represents the difference between the vertical coordinte $z$ along the column and the coordinate of the bottom of the stratified layer $z_{layer}$.
	The vertical dashed line marks the position of the transition between the neutral core and the stably stratified layer.
	For each $m$, the modes have approximately the same shape in the neutral core.
	The bumps in the energy at the right-hand side of the curves show the effect of the Ekman pumping.
	The curves are normalized by the value of their energy in the equatorial plane (not shown).}
	\label{Fig_Penetration_Profondeur_KE}
\end{figure}

\subsection{Penetration into the stratified layer}

In order to quantify the penetration of the quasi-geostrophic modes into the stratified layer, we compute profiles at a given distance $s$ from the rotation axis.
Figure \ref{Fig_Penetration_Profondeur_KE} shows the profiles of the azimuthal average of the quasi-geostrophic kinetic energy $u_s^2 + u_\phi^2$ along the axial direction, for three values of $N_0$ and two wave numbers $m$.
For a given $N_0$, the modes have approximately the same frequency.
The points are chosen along the column located at the cylindrical radius $s=0.7$, where the energy is non-zero in the equatorial plane.
The sharp bump before the decrease of the kinetic energy, observed on all curves, corresponds to the Ekman layer.
When $N_0=0.1 \, \Omega$, the Taylor column penetrates into the outer layer.
The flow just below the Ekman layer is nearly the same as that in the neutral core.
When $N_0 = \Omega$, the kinetic energy decreases in the layer, although it does not vanish before the Ekman layer is reached.
When $N_0 = 2 \, \Omega$, the kinetic energy is further reduced, and even more so for higher azimuthal modes $m$.


\begin{figure}
	\centering
	\includegraphics[width=0.90\textwidth]{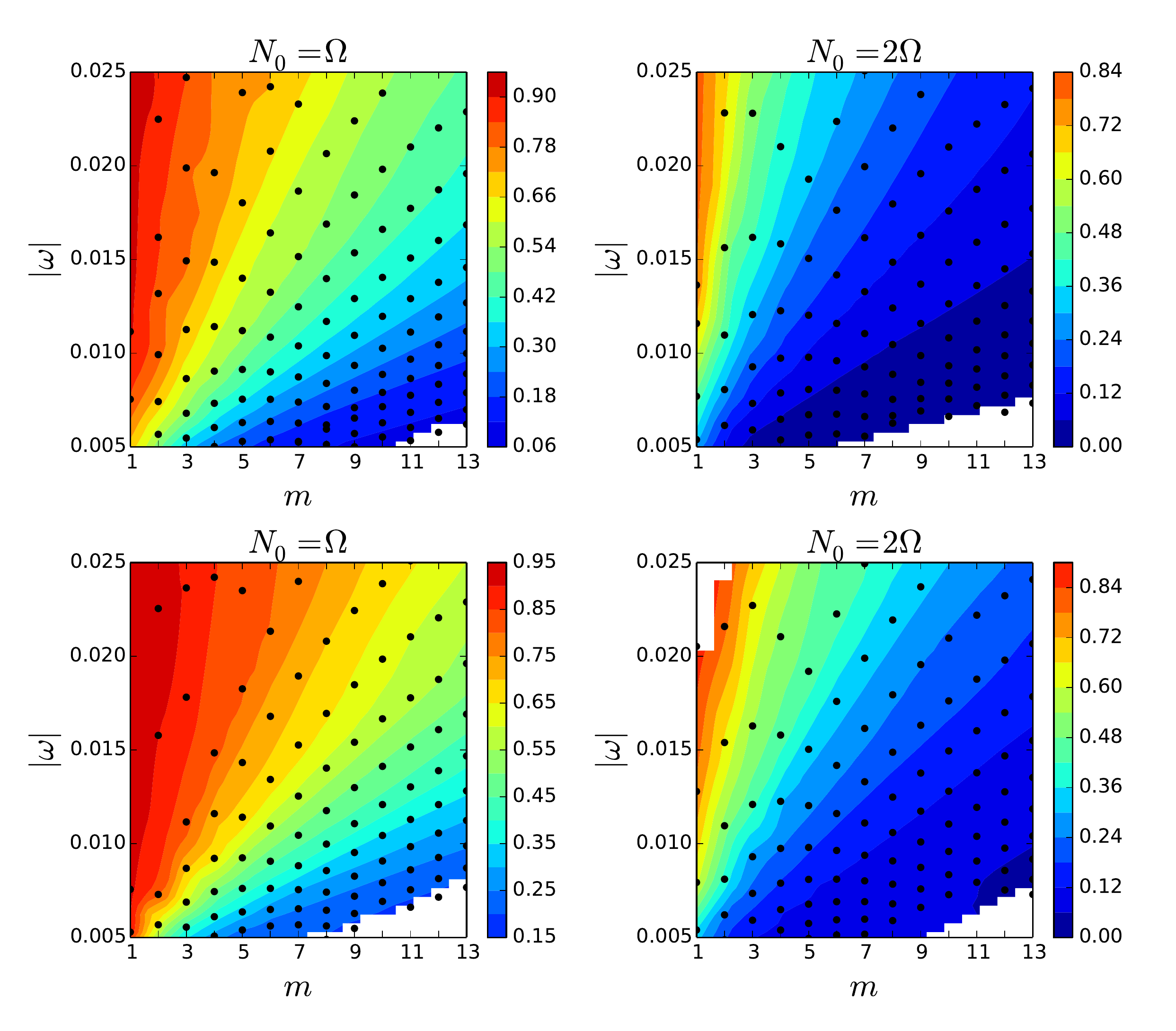}
	\caption{Maps of the amplitude of the transmission coefficient $T$ as a function of the azimuthal number $m$ and the dimensionless frequency $|\omega|$ of the modes.
	Top row: layer with constant Brunt-Väisälä frequency (eq. \ref{Eq_Model_Takehiro}); bottom row: layer with linear Brunt-Väisälä frequency (eq. \ref{Eq_Model_Linear_Profile}).
	The Brunt-Väisälä frequency is set to $N_0=\Omega$ on the left column and $N_0 = 2 \Omega$ on the right column.
	Values are computed at the cylindrical radius $s=0.7$.
	$T$ is interpolated between the black dots which represent numerical modes for which there are quasi-geostrophic columns at $s$ where the kinetic energy is non-zero.
	In the two models, the stratified layer has the same thickness (140 - 150 km).}
	\label{Fig_Penetration_T}
\end{figure}

From these profiles (Fig. \ref{Fig_Penetration_Profondeur_KE}), we can define a transmission coefficient $T$.
It is our key parameter to compare quantitatively the penetration of the flow inside the stably stratified outer layer, for eigenmodes of various shapes and frequencies.
Let us define $T$ as the ratio of the value of the square root of the azimuthal average of the quasi-geostrophic kinetic energy $u_s^2 + u_\phi^2$, taken just under the Ekman boundary layer (Figure \ref{Fig_Penetration_Profondeur_KE}), to that in the equatorial plane.
Figure \ref{Fig_Penetration_T} shows the distribution of transmission coefficients $T$ defined along the quasi-geostrophic column at a given cylindrical radius ($s=0.7$). 
Modes of different $m$ and different frequencies $\omega$ are represented.
For a given $N_0$, large-scale modes are less affected than small-scale modes which cannot penetrate deep into the layer.
This behaviour is reinforced when $N_0$ increases, showing that the stratification clearly acts as a low-pass filter: large-scale eigenmodes (high frequencies, low $m$) penetrate deeper inside the outer layer than small-scale eigenmodes (low frequencies, large $m$).

Because $N(r)$ is not known in the Earth's core, Figure \ref{Fig_Penetration_T} also compares the effect of two stratification profiles (\ref{Eq_Model_Takehiro}) and (\ref{Eq_Model_Linear_Profile}).
The modes can penetrate slightly deeper in the layer with the linear profile, but the overall distribution is similar.
Therefore, the exact shape of $N (r)$ does not seem to affect the leading order behaviour for a layer with a given thickness.
Finally, we have also observed that the thicker the layer, the less the modes penetrate into the outer layer, which is the expected behaviour \citep{Takehiro_2001strati}.

\begin{figure}
	\centering
	\includegraphics[width=0.7\textwidth]{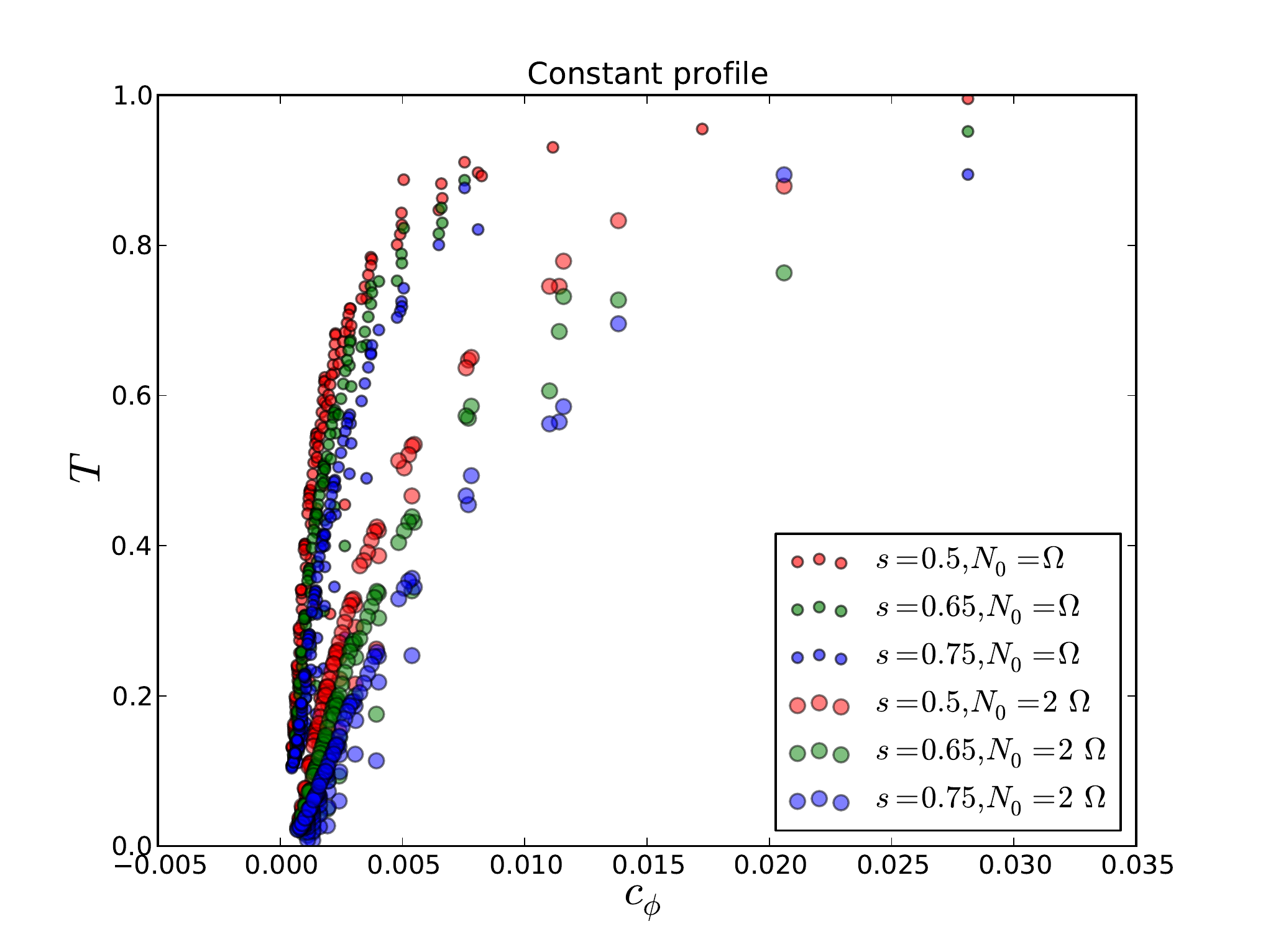}
	\caption{Dependence of transmission coefficient $T$ with the dimensionless phase velocity $c_\phi = |\omega|/m$ of the modes, for various cylindrical radius $s$ and Brunt-Väisälä frequency $N_0$.
	Computations were performed at $E=10^{-7}$, $Pr=0.1$ and $N_0 = \Omega$.
	$T$ is computed automatically for all the modes.
	Only the results for constant Brunt-Väisälä frequency (eq. \ref{Eq_Model_Takehiro}) are shown.
}
	\label{Fig_Penetration_cphi}
\end{figure}

The lines of constant transmission coefficient $T$ displayed in Figure \ref{Fig_Penetration_T} suggest that a single parameter may control $T$.
Figure \ref{Fig_Penetration_cphi} shows the transmission coefficient $T$ as a function of the dimensionless angular phase velocity $c_\phi = |\omega|/m$ for profiles taken on modes with various $m$ and $\omega$, three cylindrical radii $s$, and two values of $N_0$.
The modes with different $m$ and $\omega$ collapse quite well, although a residual dependence on $s$ (the location of the quasi-geostrophic column in the shell) and $N_0$ is now apparent.
Columns at high and mid latitudes can penetrate slightly better in the outer layer than columns at low latitudes.
At a given $s$, modes with a larger phase velocity are in general less affected by the outer layer than the others.
There are also some departures from the general trend, which are likely to come from the automatic procedure to follow modes as the parameters are changed and the automatic evaluation of $T$.

\subsection{Equatorial trapping}

\begin{figure}
	\centering
	\begin{tabular}{cc}
		\includegraphics[width=0.45\textwidth]{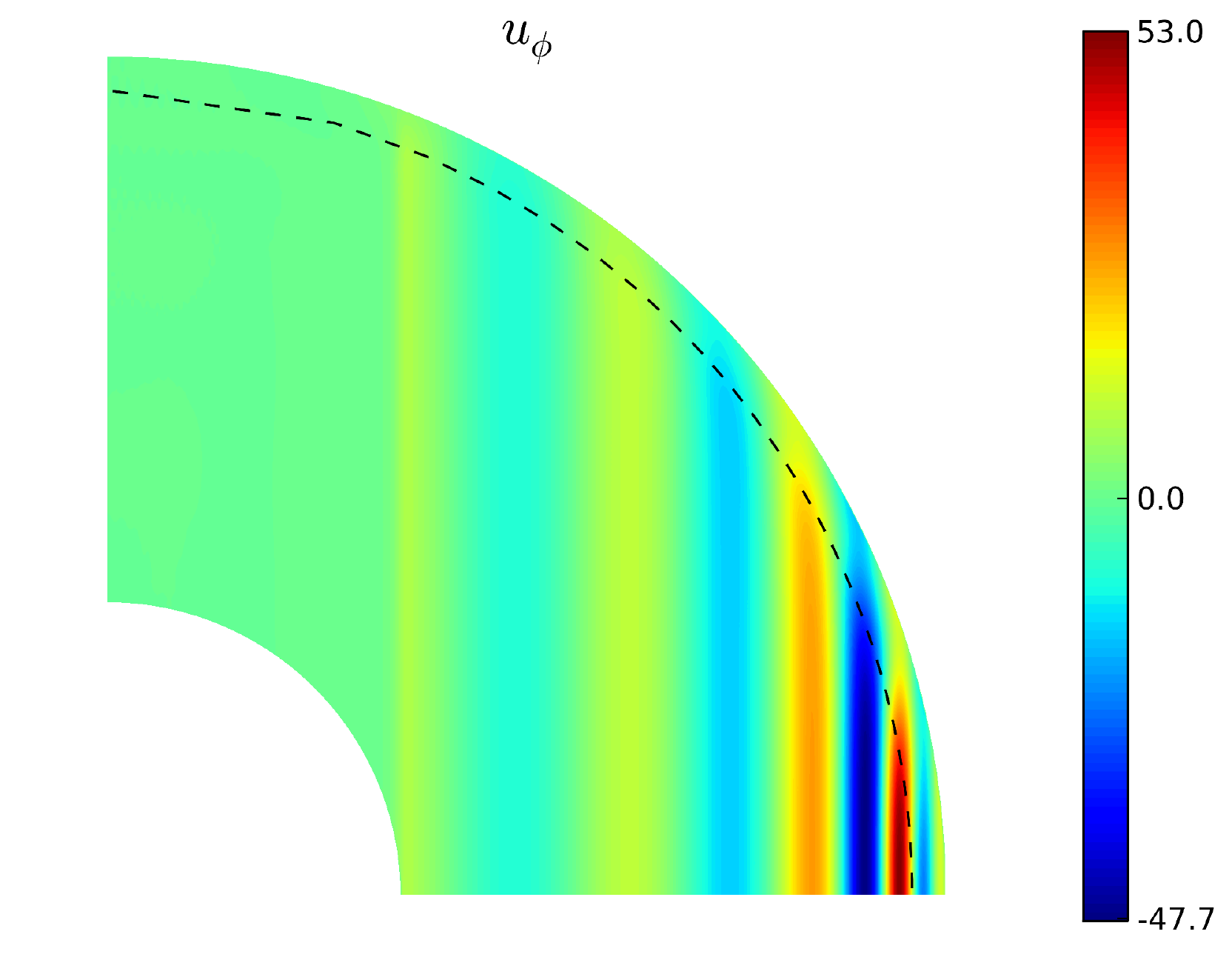} & \includegraphics[width=0.45\textwidth]{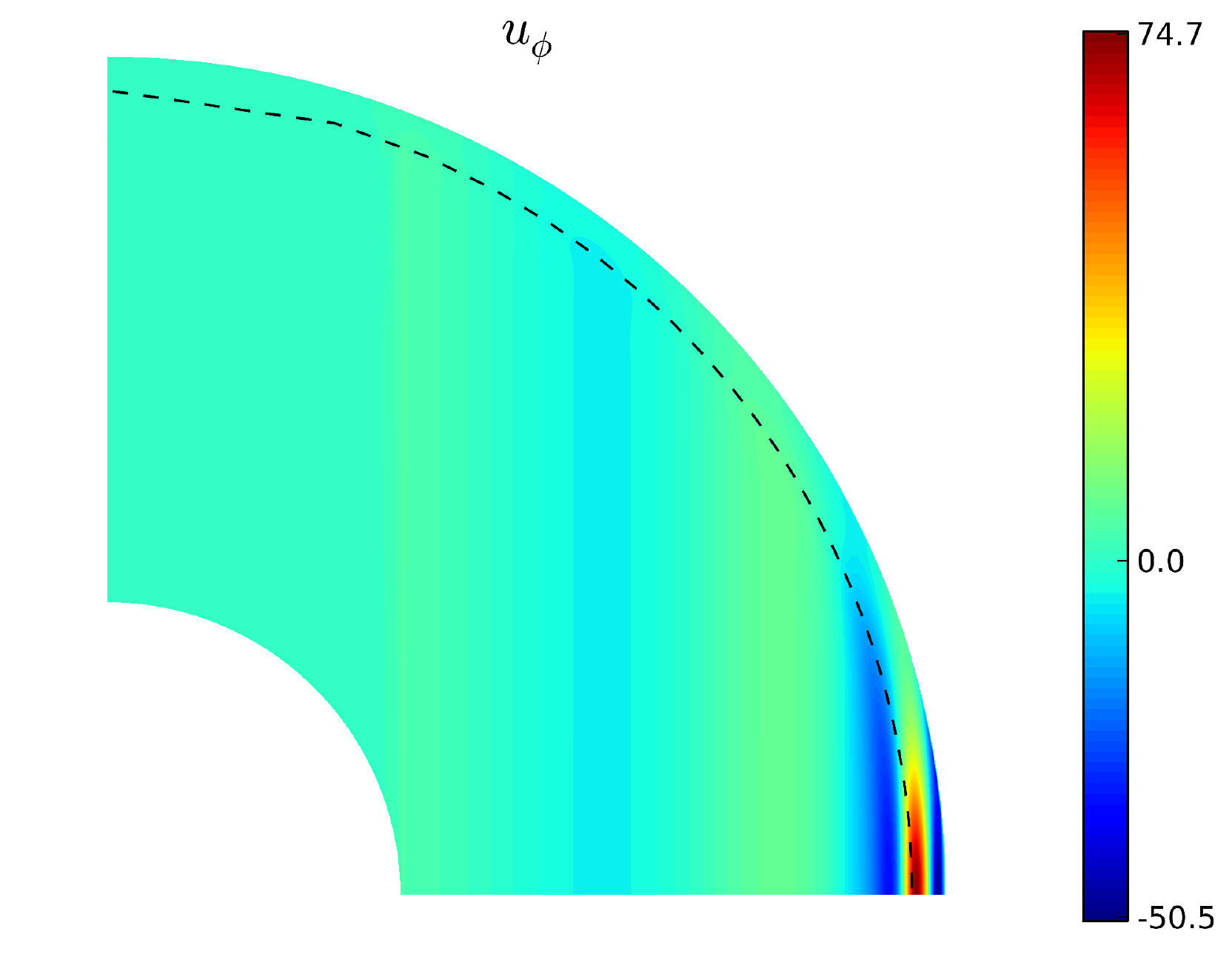} \\
		$N_0 = \Omega$ & $N_0 = 2 \Omega$ \\
		$\omega = -0.0115$ & $\omega = -0.0138$ \\
	\end{tabular}
	\caption{Distribution of the azimuthal component $u_\phi$ of the velocity field for two large-scale eigenmodes $m=1$ in a meridional plane. The dashed line represents the bottom of the stratified layer. The amplitude is arbitrary. Computations at $E=10^{-7}$ and $Pr=0.1$.}
	\label{Fig_Penetration_Trapping}
\end{figure}

A peculiar behaviour is observed near the equator.
There, we observe that modes of dimensionless frequency $|\omega| \gtrsim 0.02$ and of small wave numbers $m \leq 3$ can penetrate inside the equatorial region, where the azimuthal component $u_\phi$ can be equatorially trapped in the stably stratified layer.
This phenomenon is illustrated in Figure \ref{Fig_Penetration_Trapping}.
In this case, the modes still have $\omega < 0$, reminding of Rossby waves, but they are now governed by both Coriolis and buoyancy forces.
This trapping has been predicted by theoretical studies \citep{Friedlander_1982waves,Crossley_1984waves}.
Note however that a strong dipolar magnetic field may release the trapped modes \citep{bergman1993magnetic}.

\subsection{Thermal vs chemical stratification}
\begin{figure}
	\centering
	\includegraphics[width=0.95\textwidth]{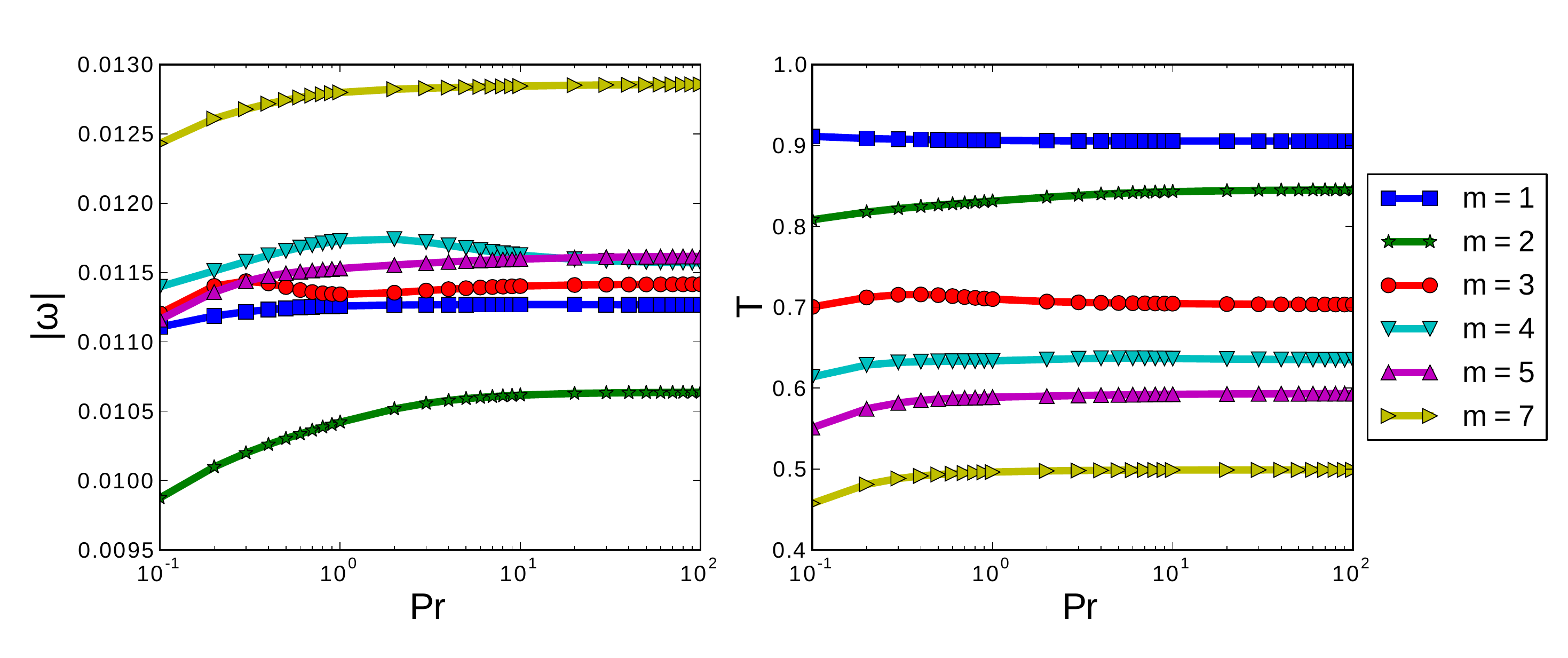}
	\caption{Evolutions of the absolute dimensionless frequency $|\omega|$ and of the transmission coefficient $T$ as a function of the Prandtl number $Pr$ of some quasi-geostrophic modes. Computations at fixed $E=10^{-7}$ and for model (\ref{Eq_Model_Takehiro}) with $N_0 = \Omega $.}
	\label{Fig_Penetration_Suivi_Pr}
\end{figure}

Thermal and chemical buoyancy are governed by the same advection-diffusion equations, the main difference being the very different value of their respective diffusivity.
Estimates of diffusivities in the Earth's core suggest that representative values of Prandtl number are $Pr \gtrsim 0.1$ for thermal diffusivity and $Pr \gtrsim 10$ for the diffusivity of light elements in the core.
We can thus study the effect of thermal or chemical stratification
by adjusting the value of the Prandtl number in our model.
Figure \ref{Fig_Penetration_Suivi_Pr} shows the dependence of the absolute dimensionless frequency $|\omega|$ and of the transmission coefficient $T$ as a function of $Pr$, for some high frequency modes of different wave numbers $m$.
For $Pr>10$, an asymptotic regime is reached where both $\omega$ and $T$ do not depend on $Pr$ any more.
In any case, $Pr$ has only a small effect on the penetration of the columns in the outer layer.




\section{Discussion}	\label{sec:discussion}

\subsection{Comparison with previous studies}

We have shown that a sufficiently stratified layer can affect the penetration of quasi-geostrophic inertial modes based on their phase speed.
A theoretical value for the penetration distance $\delta$ has been derived by \cite{Takehiro_2001strati}, using a local plane layer approximation,
\begin{equation}
\delta_{th} = \frac{2\Omega}{N_0} \, \frac{1}{\sqrt{k^2+l^2}}
\label{eq:tak_full}
\end{equation}
where $k$ and $l$ are the wavenumbers in the azimuthal and radial directions respectively.
Because $l$ is difficult to estimate, they further simplify their expression assuming isotropy $l = k = m/s$ for a global neutral convection mode of azimuthal wavenumber $m$.
Although this hypothesis is questionable (and we will discuss a better one below), they find good agreement between the penetration length obtained in numerical simulations and their estimate
\begin{equation}
\delta_{m} = \frac{2\Omega}{N_0} \, \frac{s}{\sqrt{2}\,m},
\label{eq:tak_simple}
\end{equation}
as long as $\delta \lesssim 0.4$.
However, this last expression $\delta_m$ translates into a transmission coefficient $T$ that depends only on the azimuthal wavenumber $m$, in contradiction with our findings that also exhibit a strong dependence with the frequency $\omega$ (see Fig. \ref{Fig_Penetration_T}).

We have found it difficult to measure reliably a penetration distance $\delta$ on the profiles shown on Figure \ref{Fig_Penetration_Profondeur_KE}, where the amplitude seems to decay exponentially in the lower region of the layer and then saturate to a constant value near the solid boundary.
We therefore defined a transmission coefficient, which has the drawback of being dependent of the layer thickness, but the advantage of being well defined, even for arbitrarily shaped profiles.

\begin{figure}
	\centering
	\includegraphics[width=0.8\textwidth]{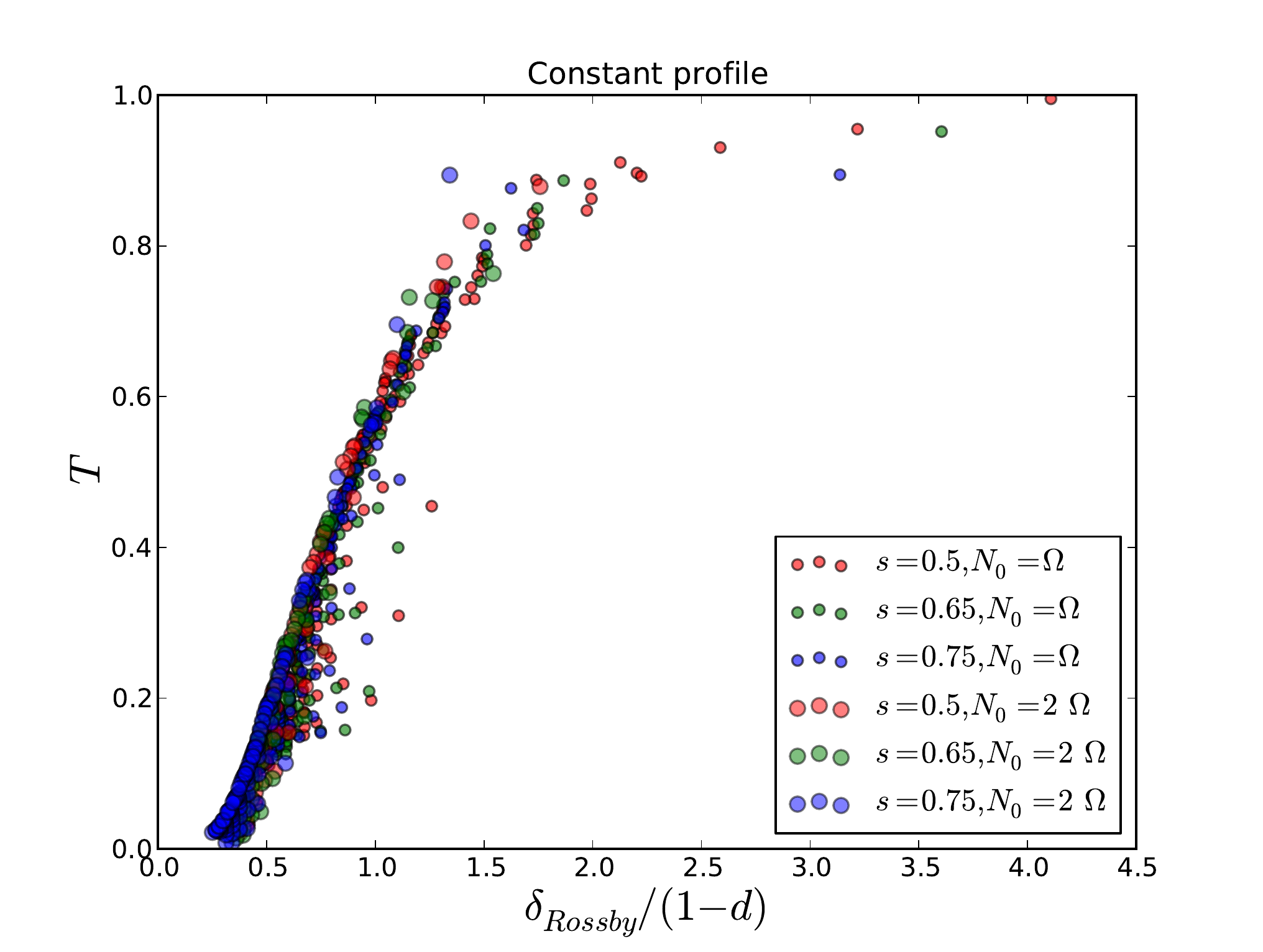}
	\includegraphics[width=0.8\textwidth]{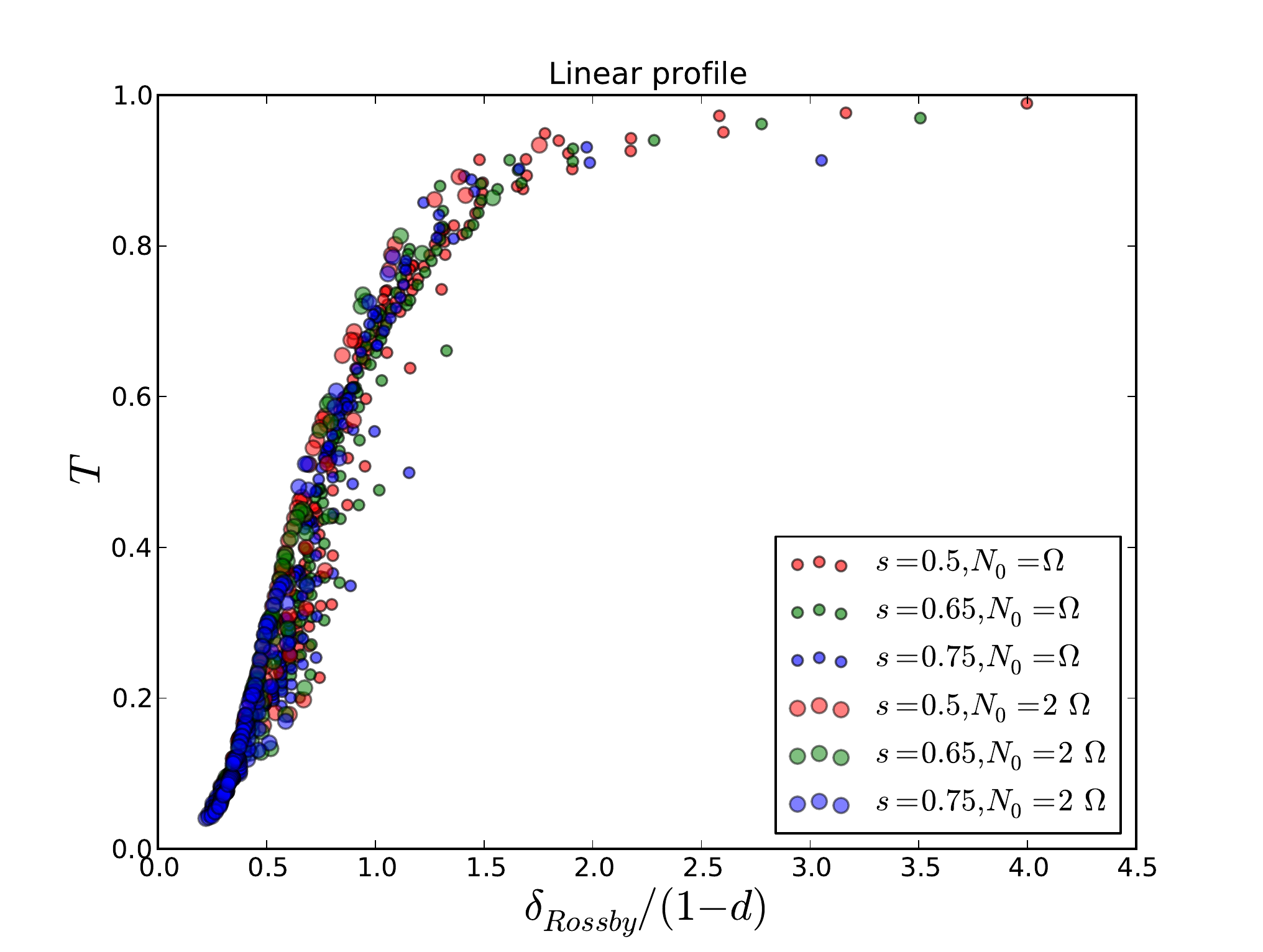}
	\caption{Transmission coefficient $T$ as a function of theoretical penetration depth $\delta_{Rossby}$ (equation \ref{eq:delta_good}), normalized by the stratified layer thickness $1-d$.
	Top: constant stratification profile (eq. \ref{Eq_Model_Takehiro}). Bottom: linear stratification profile (eq. \ref{Eq_Model_Linear_Profile}).
	All transmission coefficients obtained for various position ($s=0.5$, $s=0.65$, $s=0.75$) and two stratification strength ($N=\Omega$ and $N=2\Omega$) collapse quite well.}
	\label{Fig_final}
\end{figure}

The modes we have studied are very close to global Rossby modes, which have the following dispersion relation in a local plane approximation without stratified layer \citep{greenspan1968}:
\begin{equation}
\omega = 2\Omega \beta \frac{k}{k^2 + l^2}
\label{eq:dispersion}
\end{equation}
where $\beta = -s/(1-s^2)$ for a full sphere, and all length scales are normalized by the radius of the Earth's core $R_0$.
Ignoring the small variations of $\omega$ with $N$ (see table \ref{Table_Penetration_Eigenmodes}), we can use this expression to estimate $k^2 + l^2$ from $\omega$ and $k=m/s$.
Equation (\ref{eq:tak_full}) then becomes
\begin{equation}
\delta_{Rossby} = \frac{1}{N_0} \, \sqrt{2\Omega (1-s^2) \, \frac{|\omega|}{m}} = \frac{\cos(\theta)}{N_0}  \, \sqrt{2\Omega \, |c_\phi|},
\label{eq:delta_good}
\end{equation}
where $\theta$ is the colatitude and $c_\phi$ the phase speed in the azimuthal direction.
As shown on Figure \ref{Fig_final}, this last expression for $\delta$ that assumes the local dispersion relation of Rossby waves (eq. \ref{eq:dispersion}) seems to control all the variability in the transmission coefficient $T$.

We have thus shown that the local plane-layer theory of \cite{Takehiro_2001strati} for thermal convection extends to the whole family of Rossby waves (quasi-geostrophic inertial waves) in the core.
This implies a transmission coefficient $T$ for Rossby waves that depend only on phase speed $c_\phi=\omega/m$ and on colatitude $\theta$ in addition to the control parameters $N_0$ and $\Omega$.

\subsection{Possible observations in the geomagnetic data}

It is worth noting that the the absolute frequency of the waves increases with $N_0$ (Table \ref{Table_Penetration_Eigenmodes}).
This sensitivity of eigenmode frequencies to the Brunt-Väisäla frequency may be a way to probe for the existence of a stably stratified layer below the core-mantle boundary, but other details may also alter the frequency of the modes (ellipticity, mean flow, ...), and it will probably be difficult to tell them apart.

In addition, the relatively high frequencies of our modes are very difficult to extract from the residual noise in the current geomagnetic models.
However, the ability of the three SWARM satellites to better separate external and internal fields \citep{swarm}, may enable us to detect the signature of these modes in the geomagnetic data.

We have computed the secular variation induced by our modes acting on a model of the present geomagnetic field \citep{igrf10}.
The magnetic signatures of large-scale ($m<3$) and high frequency modes (periods shorter than two months) concentrates in the equatorial area, where they are amplified (see Figure \ref{Fig_Penetration_Suivi_Pr}).
If these waves are detected, the width of the equatorial belt in which they appear will also unveil the depth of the stratified layer.

\subsection{Influence of the magnetic field}

The quasi-geostrophic fast modes are only very weakly sensitive to the simple magnetic fields considered here (the toroidal Malkus field and the uniform axial field), for magnetic field strength in the estimated range for the Earth's core ($B_0=3\,$mT).
In particular, the magnetic field does not change their ability to penetrate the stratified layer.
It comes from the fact that these modes evolve on a time scale much faster than that of the Alfvén waves, as measured by the small Lehnert number of the Earth ($Le=10^{-4}$).
Note that the Lehnert number depends on a length-scale $\ell$: $Le(\ell) = B_0/\sqrt{\rho_0 \mu_0}\,\Omega \ell$, so that at small enough length-scales the magnetic field will affect even the fast modes.
For the Earth however, this length-scale is beyond what can be observed \citep{gillet2011}.

We expect that the fast Rossby modes remain insensitive to more complex magnetic fields.
Unfortunately, our numerical code \emph{Singe} is not well suited to take into account complex imposed fields, because they lead to an increasing number of coupled harmonic degrees, which are cumbersome to express and translate into matrix coefficients.
A possible solution would be to use computer algebra systems (CAS) to write down the coupling operators between spherical harmonics in a symbolic way \citep[as in][]{Schmitt_2010}, and then to export them into a language suitable for numerical computations.
Another more flexible approach is to use so-called matrix-free methods, which avoid to explicitly form the matrices, but require to provide appropriate preconditioners.
These methods may come forward in the forthcoming years.

\section{Conclusion}

We have studied the possible penetration of quasi-geostrophic modes in a stratified layer at the top of the Earth's core.
The ability to penetrate depends on both the frequency $|\omega|$ and the azimuthal wave-number $m$: faster modes penetrate more easily than slower modes.
Combining the finding of \cite{Takehiro_2001strati} and the dispersion relation of Rossby modes, we put forward a single parameter that accounts for all the variability in the penetration capability of the modes (eq. \ref{eq:delta_good}).

For an Earth-like magnetic field strength, the two simple field geometries we have considered here do not affect the shape or the frequency of our fast Rossby modes in any significant way.
This is due to the small Lehnert number $Le=10^{-4}$ which ensures that inertial waves are much faster than Alfvén waves at large spatial scales.
Although more complex magnetic field should be considered, it is tempting to conjecture that these fast Rossby modes are immune to any magnetic field.

Large scale modes at periods of a few months are slow enough for their magnetic signature to be detectable in principle, and can penetrate a stratified layer with $N_0 = 2\Omega$.
Close to the surface of the core, the most intense motions associated with the modes in the presence of a stratified layer are located close to the equator, where they seem to couple with a trapped mode (see Figure \ref{Fig_Penetration_Suivi_Pr}).
This is where a magnetic signature may be generated and detected.
Such a localized signature would provide observational evidence and constraints on the hypothetic stratified layer.

\paragraph{Acknowledgments}

The authors would like to thank the whole geodynamo team for fruitful discussions.
The most demanding computations were performed using the Froggy platform of the CIMENT infrastructure (\texttt{https://ciment.ujf-grenoble.fr}), supported by the Rhône-Alpes region (GRANT CPER07\_13 CIRA), the OSUG@2020 labex (reference ANR10 LABX56) and the Equip@Meso project (reference ANR-10-EQPX-29-01).
Plots were produced using matplotlib \citep[\texttt{http://matplotlib.org}]{matplotlib}.
ISTerre is part of Labex OSUG@2020 (ANR10 LABX56).

\appendix

\section{Effect of viscosity on quasi-geostrophic inertial modes}	\label{sec:visc}

Our study includes a small viscosity, which results in a non-zero damping of the modes, a small effect on their frequency, but also a significant effect on their shape, which is spiralling in the presence of sufficient viscosity.

\begin{figure}
	\centering
	\includegraphics[width=0.45\textwidth]{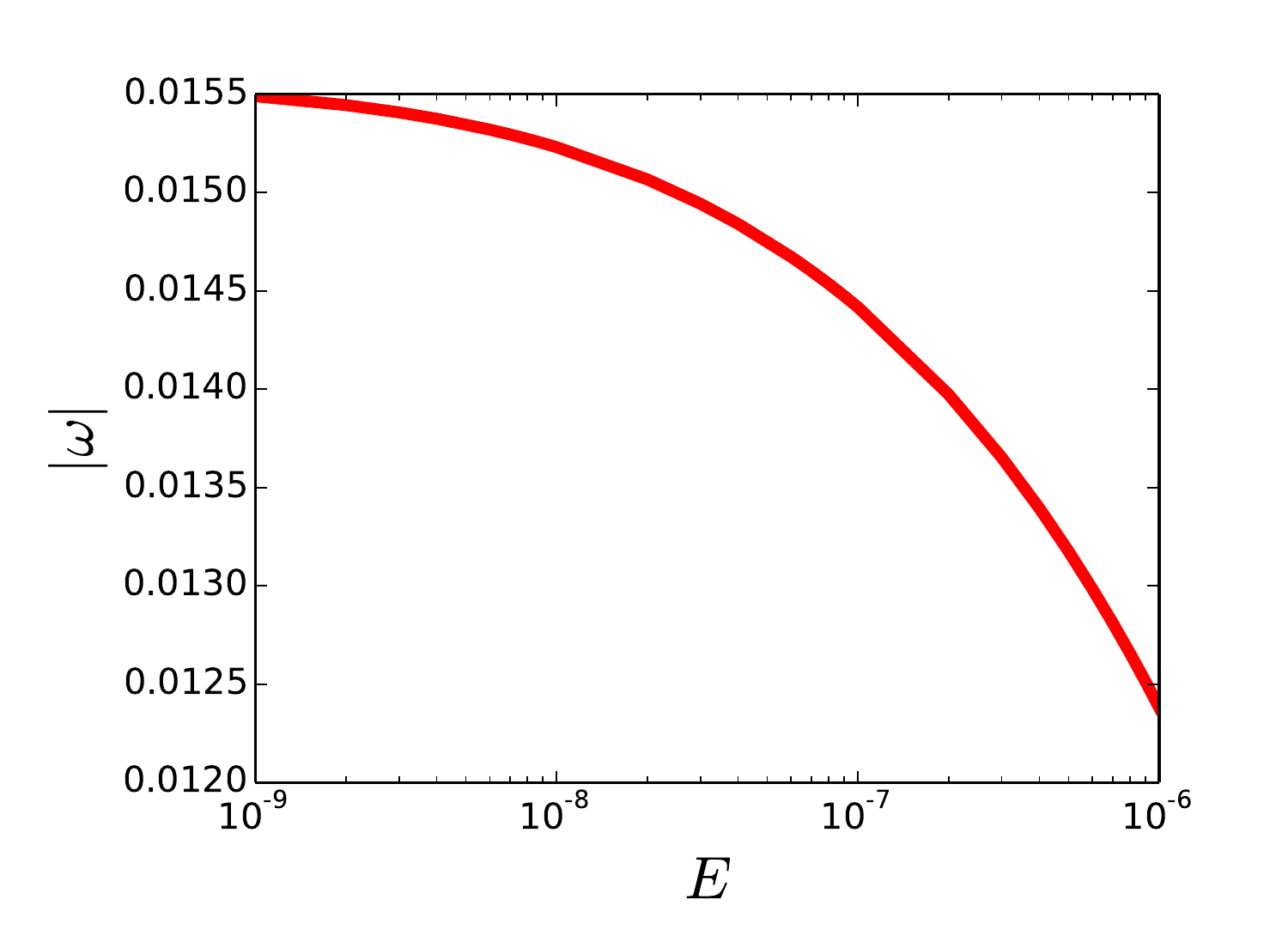}
	\includegraphics[width=0.45\textwidth]{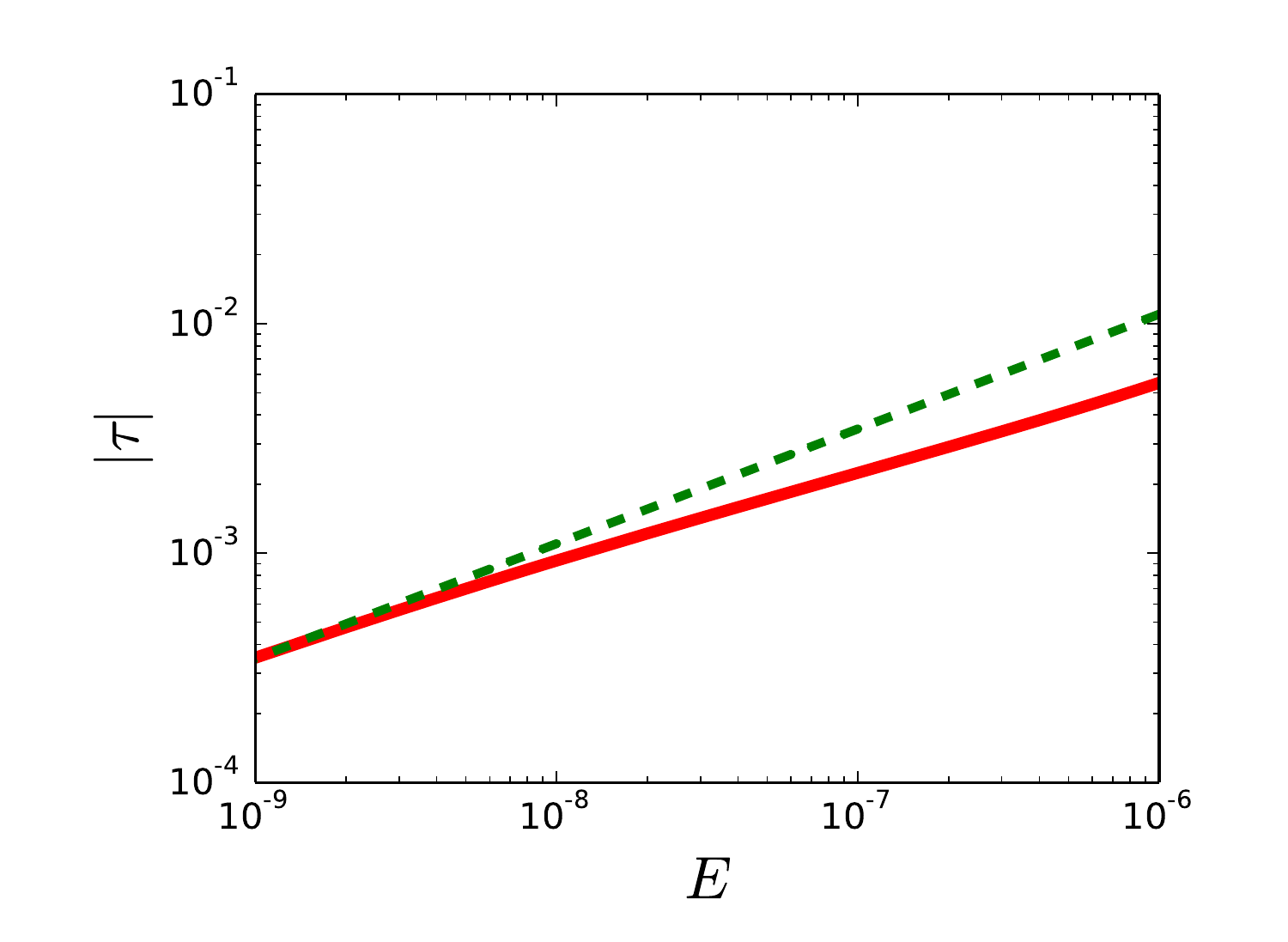}
	\caption{Frequency and damping rate of the quasi-geostrophic pure inertial mode with $m=3$, as a function of the Ekman number.
	The green dashed line correspond to an $E^{1/2}$ scaling, the expected asymptotic damping due to the Ekman layers in a full sphere \citep{liao2001viscous}.
	}
	\label{Fig_damping}
\end{figure}

\begin{figure}
	\centering
	\includegraphics[width=0.45\textwidth]{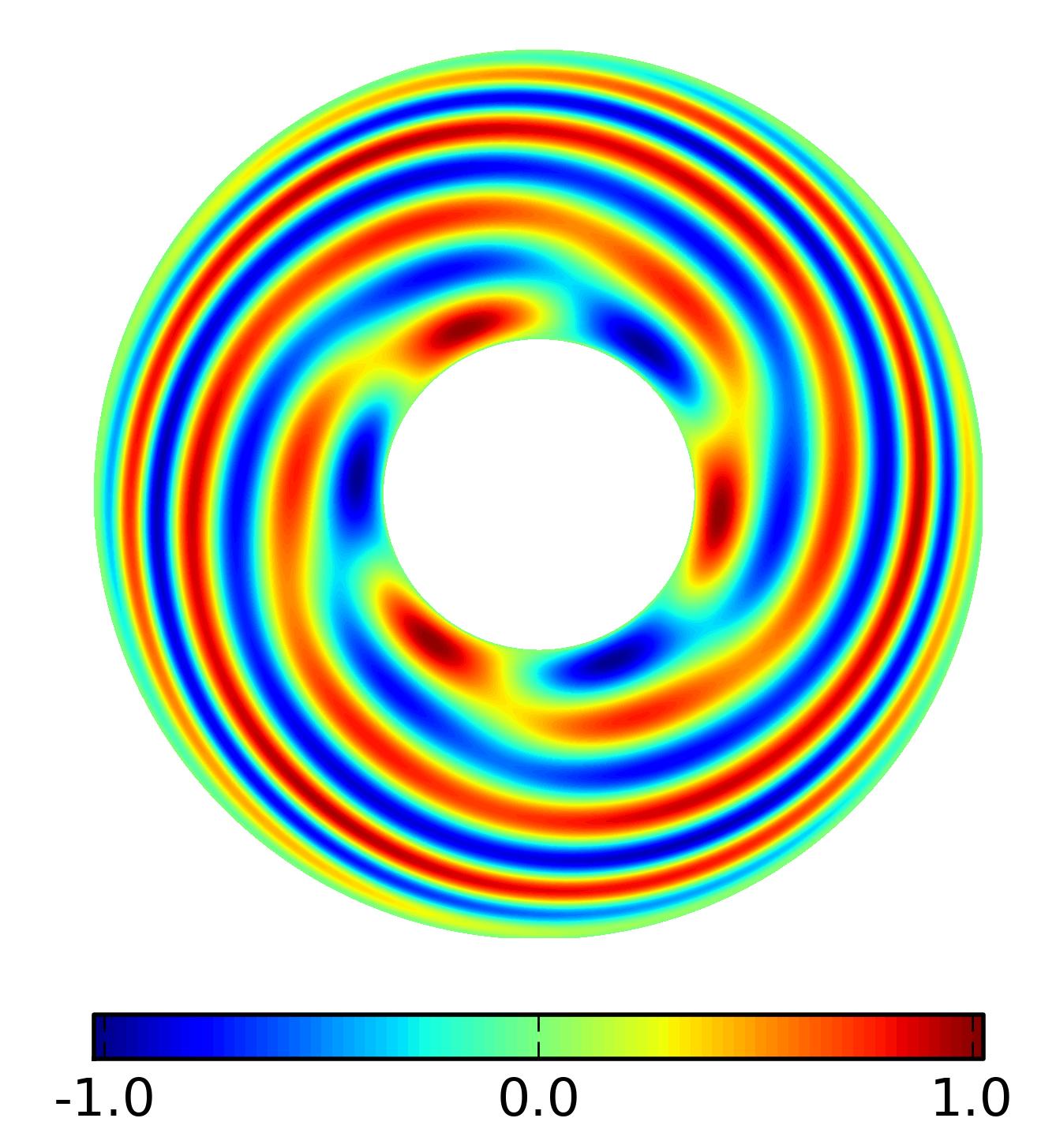}
	\includegraphics[width=0.45\textwidth]{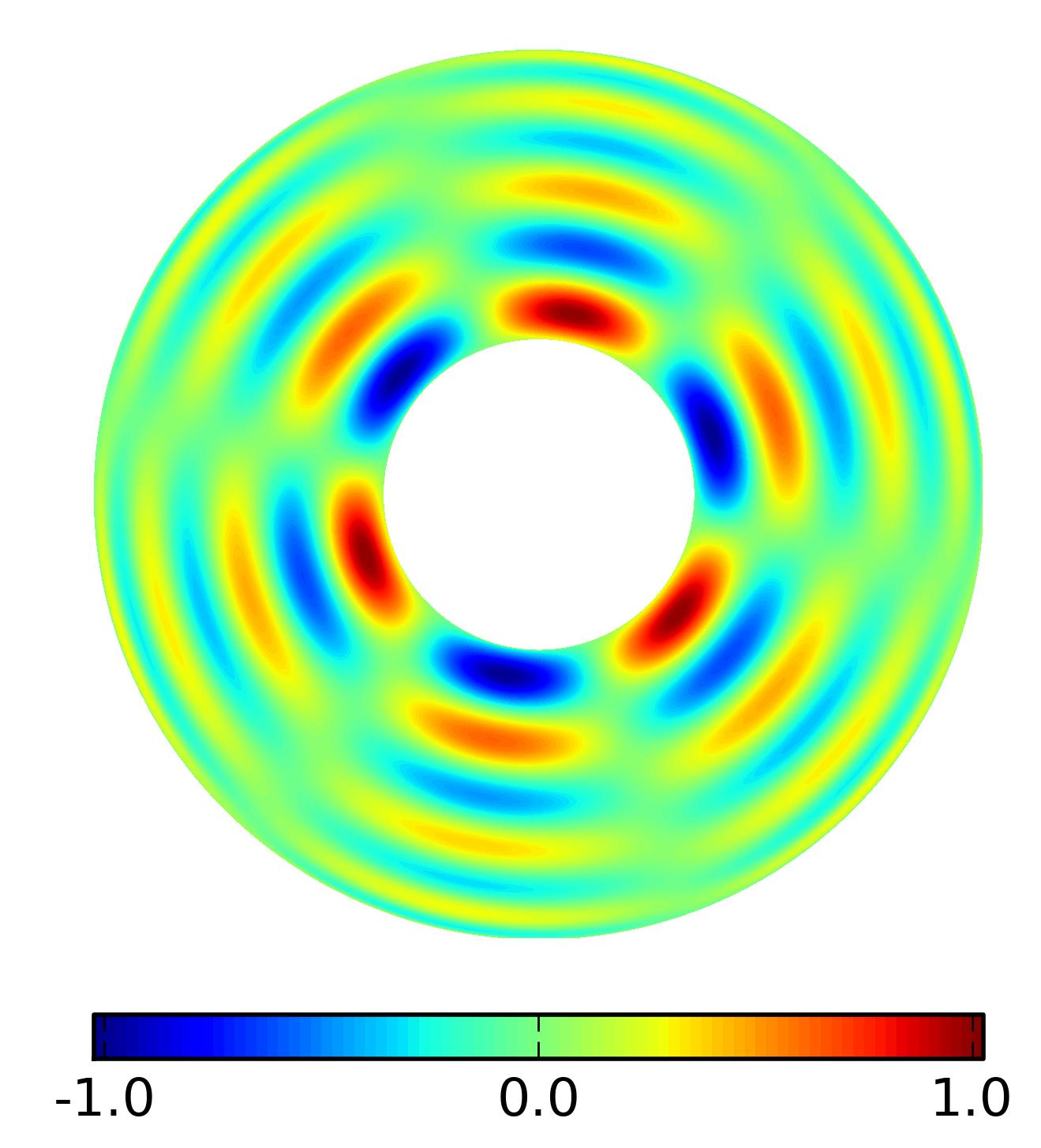}
	\caption{Equatorial cross-section of the quasi-geostrophic pure inertial mode with $m=3$ with a small viscosity. Left $E=10^{-6}$; right $E=10^{-9}$.}
	\label{Fig_spiral}
\end{figure}

As shown in Figure \ref{Fig_damping}, the asymptotic frequency and damping rate are approached only when $E \sim 10^{-9}$.
Coincidentally, Figure \ref{Fig_spiral} shows that the spiralling shape vanishes at the same very low Ekman number.

This contrasts with thermal convection in a rapidly rotating sphere, where tilted or spiralling structures are a feature of the onset \citep[e.g.][and references therein]{dormy2004,takehiro2008}.
The main reason being that, at the scale at which the convection first occurs, the viscosity is never negligible.

Note also that this does not depend on the boundary conditions (stress-free or no-slip), showing that the spirallization is an effect of the viscosity in the bulk. 

All this shows that very low Ekman numbers are needed to have an asymptotic structure of quasi-geostrophic inertial waves.

\bibliography{./biblio_qg_strati}
\bibliographystyle{plainnat}

\end{document}